\documentclass[twocolumn,hyperpdf,amsmath,amssymb,aps,prd,10pt,superscriptaddress,nofootinbib,noeprint,preprintnumbers,floatfix]{revtex4}

\usepackage{graphicx, color}
\usepackage[dvipsnames]{xcolor}
\usepackage{booktabs}
\usepackage{graphicx}

\usepackage{orcidlink}

\usepackage[letterspace=-10]{microtype} 

\usepackage{bm, amsmath, amsfonts, amssymb,xfrac}
\usepackage{multirow, tabularx, dcolumn, soul,ulem}
\usepackage{mathtools, leftidx, braket, slashed, cancel, bigdelim}
\usepackage{blkarray}
\usepackage[figures]{rotating}

\usepackage{tikz}
\usetikzlibrary[plotmarks]
\usepackage{anyfontsize}

\usepackage[utf8]{inputenc} 
\usepackage{hyperref}

\definecolor{jlab_red}{RGB}{192,39,45}
\definecolor{jlab_orange}{RGB}{249,102,0}
\definecolor{jlab_blue}{RGB}{47,122,121}
\definecolor{jlab_green}{RGB}{65,125,10}
\definecolor{jlab_gray}{gray}{0.6}
\definecolor{magenta}{rgb}{0.5, 0, 0.5}

\newcommand\bef{\begin{figure}}
\newcommand\eef[1]{\label{fg:#1}\end{figure}}
\newcommand\beq{\begin{equation}}
\newcommand\eeq[1]{\label{#1}\end{equation}}
\newcommand\beqa{\begin{eqnarray}}
\newcommand\eeqa[1]{\label{#1}\end{eqnarray}}
\newcommand\bet{\begin{table}}
\newcommand\eet[1]{\label{tb:#1}\end{table}}

\newcommand{\be}{\begin{equation}}
\newcommand{\ee}{\end{equation}}
\newcommand{\bea}{\begin{eqnarray}}
\newcommand{\eea}{\end{eqnarray}}
\newcommand{\beas}{\begin{eqnarray*}}
\newcommand{\eeas}{\end{eqnarray*}}

\newcommand{\Tr}{\textrm{Tr}}

\newcommand{\rub}{\affiliation{Institut f\"ur Theoretische Physik II, Ruhr-Universit\"at Bochum, D-44780 Bochum, Germany }}

\hypersetup{%
pdftitle = {title},
pdfsubject = {},
pdfkeywords = {},
colorlinks = {true},
filecolor = {black},
linkcolor = {jlab_blue},
menucolor = {black},
citecolor = {jlab_green},
urlcolor = {jlab_green},
}{}

\begin{document}

\title{Shallow $T_{bc}$ states from an EFT analysis of $B^{(*)} \bar D^{(*)}$ scattering on the lattice}

\begin{abstract}

We present an effective field theory (EFT) framework for coupled-channel $B^{(*)}\bar D^{(*)}$  scattering, applying it to recent lattice QCD results by Alexandrou et al.~[Phys.~Rev.~Lett.~132, 151902 (2024)]. Two complementary EFT approaches are developed:
(1) A low-energy theory near the $B \bar D$ ($J=0$) and $B^* \bar D$ ($J=1$) thresholds, where coupled-channel effects are integrated out;
(2) A coupled-channel formulation, where all relevant momentum scales are treated as soft, incorporating contact interactions and one-pion exchange (OPE). Importantly, OPE contributes to the lowest channels only through off-diagonal transitions, thus resulting in the appearance of the  left-hand cut from two-pion exchange. 
The two approaches yield mutually consistent results, supporting the existence of shallow bound states in both channels, in agreement with the lattice findings. 
The finite-volume spectra and extracted pole positions show a near-degeneracy in $J=0$ and $J=1$ channels, consistent with heavy-quark spin 
symmetry (HQSS). Using HQSS, we predict additional shallow bound states near the $B \bar D^*$ and $B^* \bar D^*$ thresholds, which are accessible to future lattice simulations. The effect of OPE on the finite volume spectra is found to be small, with only moderate impact on HQSS partners.

\end{abstract}

\author{Michael Abolnikov \orcidlink{0009-0006-9874-0876}}
 \rub

\author{Lu Meng \orcidlink{0000-0001-9791-7138}}
\rub\affiliation{School of Physics, Southeast University, Nanjing 211189, China }

\author{Vadim Baru\orcidlink{0000-0001-6472-1008}}
\rub

\author{Evgeny Epelbaum\orcidlink{0000-0002-7613-0210}}
\rub

\author{Arseniy~A.~Filin\orcidlink{0000-0002-7603-451X}}
\rub

\author{Ashot~M.~Gasparyan\orcidlink{0000-0001-8709-4537}}
\rub

\maketitle

\section{Introduction}

Among the most compelling discoveries in the field of exotic hadrons is the recent experimental observation of the $T_{cc}$ state by the LHCb collaboration~\cite{lhcb:2021vvq,lhcb:2021auc}. The discovery of this manifestly exotic isoscalar state with minimal quark content $cc\bar{u}\bar{d}$ marks a significant milestone, given that the $T_{cc}$ is the first experimentally confirmed doubly-charm tetraquark. Its remarkably narrow width and mass, lying just below the lowest meson-meson threshold ($D^{*+}D^0$), strongly suggest a molecular nature.
The isoscalar $D D^*$ scattering channel has also been studied in lattice QCD~\cite{Padmanath:2022cvl,Chen:2022vpo,Whyte:2024ihh,Collins:2024sfi,Lyu:2023xro}, albeit at unphysically large pion masses. These studies confirm the presence of a near-threshold state. In contrast, no indication of a shallow near-threshold state has been observed in the isovector  $D D^*$ channel, neither experimentally~\cite{lhcb:2021vvq,lhcb:2021auc} nor in lattice QCD simulations~\cite{Meng:2024kkp,PitangaLachini:2025pxr}.

Apart from the $T_{cc}$, its heavier doubly-heavy analogs have also attracted significant interest:
the doubly-bottom tetraquark $T_{bb} \sim bb\bar{q}_1\bar{q}_2$, and the mixed bottom-charm system $T_{bc} \sim bc\bar{q}_1\bar{q}_2$ as well as their antiparticles. Lattice QCD studies have converged on the existence of a deeply bound $T_{bb}$ state with quantum numbers $I(J^P) = 0(1^+)$, which is predicted to be stable under both strong and electromagnetic interactions, decaying only via the weak force~\cite{Francis:2016hui,Junnarkar:2018twb,Leskovec:2019ioa,Mohanta:2020eed,Meinel:2022lzo,Hudspith:2023loy,Alexandrou:2024iwi}. 
In contrast, the situation for $T_{bc}$ remains much less clear. 
In what follows, we collectively denote the tetraquark states with quark content $bc\bar q_1\bar q_2$ and their antiparticles $\bar b\bar c q_1 q_2$ by $T_{bc}$.
With one bottom and one charm quark, $T_{bc}$ occupies a unique position between the  $T_{cc}$ and $T_{bb}$, offering an important testing ground for the interplay of heavy-quark symmetries and low-energy QCD dynamics.

Initial lattice hints of the existence of a bound $T_{bc}$ tetraquark with $I(J^P) = 0(1^+)$ were reported in~\cite{Francis:2018jyb}, but subsequent calculations with improved statistics and larger volumes did not confirm this result~\cite{Hudspith:2020tdf,Meinel:2022lzo}. More recently, a Lüscher-based analysis of ground-state energy levels at various light-quark masses (corresponding to pseudoscalar meson masses ranging from 0.5 to 3.0 GeV) in Ref.~\cite{Padmanath:2023rdu} revealed an attractive interaction in the $I(J^P) = 0(1^+)$ $\bar{B}^* D$ channel. Extrapolation to the physical pion mass and the continuum limit yielded a bound state  $43^{+7}_{-6}{}^{+24}_{-14}$ MeV below threshold. A similar analysis was performed for the $I(J^P) = 0(0^+)$ $\bar{B} D$ channel, leading to analogous conclusions~\cite{Radhakrishnan:2024ihu}. The $\bar b \bar c u d$ four-quark system has also been studied using the Born–Oppenheimer approximation in Ref.~\cite{Hoffmann:2025frk}, employing the antistatic-antistatic potentials computed with lattice QCD in Ref.~\cite{Bicudo:2015kna} at
three values of the pion mass between 340 and 650 MeV, and extrapolated to the physical pion mass. This approach leads to the appearance of virtual $T_{bc}$ states located quite far from the lowest meson–meson thresholds for quantum numbers $I(J^P) = 0(0^+)$ as well as $0(1^+)$.

Another recent lattice QCD study~\cite{Alexandrou:2023cqg} performed a Lüscher-based analysis of both ground and multiple excited states in the $I(J^P) = 0(0^+)$ $B \bar{D}$ and $I(J^P) = 0(1^+)$ $B^* \bar{D}$ channels, aiming to determine the energy dependence of the scattering amplitudes near their respective thresholds. Using lattice QCD results at the pion mass of approximately 220 MeV and lattice spacing of 0.12 fm and utilizing the effective range expansion (ERE), this study identified shallow bound states a few MeV below both $B \bar{D}$ and $B^* \bar{D}$ thresholds. In addition, broad resonance poles above thresholds were found in both channels.

The growing number of more sophisticated lattice calculations yielding energy spectra beyond the ground state reinforces the need for equally advanced theoretical methods to reliably extract key physical quantities, such as infinite-volume (IFV) scattering amplitudes and pole positions, from finite-volume (FV) energies.
In particular, Refs.~\cite{Du:2023hlu,Baru:2016evv,Baru:2015ira}  emphasized that the ERE is only valid for parameterizing the near-threshold energy behavior of the inverse scattering amplitude in the absence of nearby left-hand cuts (lhc) that can arise from long-range interactions. The validity of the Lüscher formula~\cite{Luscher:1986pf,Luscher:1990ux,Kim:2005gf,Briceno:2014oea}, the underlying method for extracting IFV amplitudes from FV energy levels, can be called into  question in the presence of a nearby lhc. Consequently, several extensions of the Lüscher method and alternative approaches have been proposed to address this issue~\cite{Raposo:2023oru,Meng:2023bmz,Bubna:2024izx,Hansen:2024ffk, Dawid:2024oey,Yu:2025gzg}.

In particular, a chiral EFT based method for two heavy mesons consistent with unitarity and analyticity was proposed in Ref.~\cite{Meng:2023bmz}. This framework was successfully applied to analyze $DD^*$ scattering in the isoscalar ($I=0$)~\cite{Meng:2023bmz} and isovector ($I=1$)~\cite{Meng:2024kkp} channels at a pion mass of approximately 280 MeV. In this approach, FV spectra are computed by solving the Lippmann–Schwinger equation, with the  
effective potential involving both the long-range one-pion exchange (OPE) and short-range interactions parameterized by contact terms.
The  relevant low-energy constants are extracted by fitting the FV spectra, and the  resulting $T_{cc}$ pole 
in the isoscalar  $DD^*$ channel is identified as a near-threshold resonance in the corresponding infinite-volume scattering amplitude. Notably, Ref.~\cite{Meng:2023bmz} showed that the OPE (i) generates sizable repulsion  in the isoscalar channel,  
(ii) governs the energy dependence of the $S$- and $P$-wave amplitudes near threshold---thereby contributing to range corrections---and (iii) controls the leading exponentially suppressed FV corrections that are neglected in the Lüscher formalism.

Furthermore, the OPE turns out to have nontrivial impact on the light-quark mass dependence of the $T_{cc}$ pole~\cite{Abolnikov:2024key}. While its effect is minor at pion masses just above the physical value, in agreement with the HAL QCD results~\cite{Lyu:2023xro}, it becomes increasingly important at larger pion masses, eventually shifting the pole into the complex energy plane.

In contrast, the OPE’s influence on $DD^*$ scattering in the isovector channel is significantly weaker, consistent with the  
suppression of isospin coefficients relative to the isoscalar case~\cite{Meng:2024kkp}. The main difference between the lattice results in the isoscalar and isovector channels at $m_\pi \simeq 280$ MeV was attributed  to the exchange of isovector-vector mesons, associated either with the $\rho$-meson or with two-pion exchange (TPE)  mechanisms.
This interpretation is further supported by the results of Ref.~\cite{Chacko:2024cax}, where the leading TPE contributions were calculated at next-to-leading order (NLO) in chiral EFT with coupled channels. Indeed, these TPE loops were found to be strongly attractive in the isoscalar channel but repulsive in the isovector channel. Nonetheless, while the two pion exchange was found to be important, 
it was also shown that  these potentials can be largely absorbed into the  NLO contact interactions, with minor residual nonanalytic TPE contributions.

In this work, we  
apply EFT to analyze lattice data for $B^{(*)}\bar D^{(*)}$ scattering in the context of $T_{bc}$ states~\cite{Alexandrou:2023cqg}. To this end, we employ two approaches:

1. In the first approach, momentum scales associated with coupled-channel dynamics are treated as hard and integrated out, leading to a low-energy EFT valid near the lowest thresholds ($B \bar D$ for $J=0$, $B^* \bar D$ for $J=1$). 
In this regime,  the OPE does not contribute to elastic interactions, and pion dynamics is encoded in contact terms only.
Thus, a purely contact  theory is employed, with low-energy constants fitted to the lowest-lying finite-volume energy levels, up to momenta $k_{\text{max}} \simeq 310$–$380$ MeV.

2.  In the second approach, coupled-channel momentum scales are treated as soft, leading to the formulation which includes both contact interactions and the OPE, with the OPE contributing to the lowest channels only through off-diagonal transitions. It allows us to perform a coupled-channel EFT analysis that incorporates TPE effects—contributing to the lowest-lying channels through loops with intermediate states from higher thresholds—and their associated left-hand cuts. All relevant lattice data from the lowest to higher thresholds are included in the fits.

As detailed below, the two approaches yield  consistent results within uncertainties and support the existence of shallow bound states in both lowest channels, in agreement with the findings of Ref.~\cite{Alexandrou:2023cqg}. 
The nearly identical spectra in the $J=0$ and $J=1$ channels align with heavy-quark spin symmetry (HQSS), suggesting nearly identical $B \bar D \to B \bar D$ and $B^* \bar D \to B^* \bar D$ interactions. HQSS also predicts shallow states near $B \bar D^*$ and $B^* \bar D^*$ thresholds, which can be searched for in future lattice QCD simulations. TPE effects are found to be negligible, but the effect of the OPE on the shallow states near the $B^* \bar D^*$ thresholds is more pronounced. Off-diagonal contact terms are found to be strongly correlated with diagonal ones, yielding no significant improvement in the fit quality.

Finally, we note that 
theoretical implications of HQSS, including the possible existence of spin partners for the $B\bar D$ states, have been discussed in the literature (see, e.g., Refs.~\cite{Li:2012ss, Sakai:2017avl,Abreu:2022sra, Meng:2023jqk, Liu:2025fhl}). However, in the absence of experimental data, no reliable methods for constraining the parameters accompanying the short-range interactions were available until the recent lattice QCD results  of Ref.~\cite{Alexandrou:2023cqg} were recently reported.
We also emphasize that the above-mentioned predictions for $B^{(*)}\bar D^{(*)}$ states based on heavy-flavor symmetry are afflicted with uncontrolled uncertainties, since it is not possible to relate different heavy-flavor sectors containing two heavy quarks within a systematic EFT framework~\cite{Baru:2018qkb}.

In Sec.~\ref{Sec:frame}, we briefly outline the formalism employed in our calculation. Section~\ref{Sec:EFT1} presents the results of our analysis of the lattice data near the $B\bar D$ and $B\bar D^*$ thresholds within a single-channel EFT. In Sec.~\ref{Sec:EFT2}, we extend the analysis to a broader energy range, examining the role of coupled-channel off-diagonal interactions, the OPE, and higher-order terms. Our conclusions are summarized in Sec.~\ref{Sec:summary}. Additional details are provided in the Appendices. Specifically, Appendix~\ref{QQapp} shows quantile-quantile (QQ) plots supporting our statistical analysis;  Appendix~\ref{appendix_B} focuses on the details of the fitting procedure and statistical uncertainty estimates; and Appendix~\ref{App:EFT2_results}
presents consistency checks and additional uncertainty estimates.

\section{Framework}
\label{Sec:frame}

Since the lattice QCD data for FV energy levels of Ref.~\cite{Alexandrou:2023cqg} are located near the relevant thresholds, it is natural to analyze them using an EFT approach.  This allows systematic calculations of the IFV observables with reliable error estimates.  
A similar EFT approach was used by some of the authors to analyze lattice data for $DD^*$ scattering in the isospin-0 and isospin-1 channels in the context of doubly-charmed tetraquark states~\cite{Meng:2023bmz,Meng:2024kkp}. The present work describes the first application of a coupled-channel extension of this EFT approach to lattice QCD data.

Below, we outline the general aspects of the employed EFT framework. Specific details of the two approaches used to analyze lattice data for $B^{(*)}\bar D^{(*)}$ scattering are presented in Sections~\ref{Sec:EFT1} and~\ref{Sec:EFT2}. 
 
To describe the interaction between heavy mesons in an EFT framework, we employ a low-energy expansion of the effective potential in powers of $Q$,
where $Q$ denotes a typical soft scale in the problem, assumed to be well separated from the hard scales.
Furthermore, as $B$ and $D$ mesons contain heavy quarks, $ B^{(*)}\bar D^{(*)} $ interactions are expected to be constrained by HQSS, which dictates that interactions depending on the spins of heavy quarks are suppressed by the heavy-quark mass. Then, the leading-order (LO) $ B^{(*)}\bar D^{(*)} $ S-wave contact interaction follows from the effective Lagrangian, which for isospin-0 can be written as:
\be\label{Lag_LO_CT}
\begin{split}
{\cal L}_{HH}^{\rm LO} = & \frac{C_{00}}{2} \Tr \bigl( {\bar{H}_D^\dagger H_B^\dagger H_B \bar{H}_D} \bigr) \\ 
+ &\frac{C_{01}}{2} \Tr \bigl( {\bar{H}_D^\dagger \sigma_i H_B^\dagger H_B  \sigma_i\bar{H}_D} \bigr).
\end{split}
\ee
Here, $C_{00}$ and $C_{01}$ are two low-energy constants (LECs) accompanying   contact interactions between the heavy-light mesons grouped into the superfields
\bea\nonumber
H_B &=& B + \bm B^* \cdot \bm \sigma,\\
{\bar H}_D &=& {\bar D} -  { \bm \bar  {\bm D}}^* \cdot \bm \sigma,
\eea
where $B (D)$ and $B^* (D^*)$ are pseudoscalar and vector meson fields. 
Then, using the following basis states for different $J^P$ quantum numbers (written in the spectroscopic notation $^{2S+1}L_J$ with $S, L$ and $J$ referring to the spin, orbital angular momentum and total angular momentum, respectively),
\begin{equation}\label{Eq:JP}
\begin{split}
& 0^{+}: \left\{ {B}\bar D(^1S_0), {B}^*\bar D^*(^1S_0) \right\}, \\
& 1^{+}: \left\{ {B}^*\bar D (^3S_1), {B}\bar D^* (^3S_1), {B}^*\bar D^*(^3S_1) \right\}, \\
& 2^{+}: \left\{ {B}^*\bar D^*(^5S_2) \right\},
\end{split}
\end{equation}
the LO coupled-channel S-wave potentials read   
\bea
\label{vfull0+}
\hspace{-0.5cm}V_{\rm LO}^{\rm CT}[0^{+}] 
&=&\left(\begin{array}{cc}
\mathcal{C}_{d}+\mathcal{C}_{f} \hspace{0.3cm} 
& \sqrt{3}\, \mathcal{C}_{f}\\
 \sqrt{3}\, \mathcal{C}_{f} \hspace{0.3cm} & \mathcal{C}_{d}-\mathcal{C}_{f}\hspace{0.2cm}   
\end{array}\right),\\\nonumber \\
\label{vfull1+}
\hspace{-0.5cm}V_{\rm LO}^{\rm CT}[1^{+}] 
&=&\left(\begin{array}{ccc}
\mathcal{C}_{d}+\mathcal{C}_{f} \hspace{0.3cm} 
& \mathcal{C}_{f}  \hspace{0.3cm} & \sqrt{2}\, \mathcal{C}_{f}\\
\mathcal{C}_{f}  \hspace{0.3cm} & \mathcal{C}_{d} +\mathcal{C}_{f}  \hspace{0.2cm}   \hspace{0.3cm} 
& -\sqrt{2}\, \mathcal{C}_{f}  \\
\sqrt{2}\,\mathcal{C}_{f}  \hspace{0.3cm} &-\sqrt{2}\, \mathcal{C}_{f} \hspace{0.2cm}   \hspace{0.3cm} 
&  \mathcal{C}_{d}
\end{array}\right),\\
\label{vfull2+}
\hspace{-0.5cm}
V_{\rm LO}^{\rm CT}[2^{+}] 
&=&\  \mathcal{C}_{d}+2\, \mathcal{C}_{f},
\eea
where we introduced the redefined LECs $\{ \mathcal{C}_{d}, \mathcal{C}_{f}\}$ according to   $\mathcal{C}_{d}=2 (C_{00}+C_{01})$ and $\mathcal{C}_{f}= C_{01}-C_{00}$.

In addition to the LO contact potentials from Eqs.~\eqref{vfull0+}-\eqref{vfull2+},   the one-pion exchange also contributes at LO.  
However, the three-pseudoscalar-meson vertex, $B \bar D \pi$, is forbidden by angular momentum and parity conservation. As a result, the OPE is only allowed in elastic
  $B^* \bar D^*$ interactions and in coupled-channel transitions $J^P=0^+$
  $B \bar D \to B^* \bar D^*$   
  and $1^+$ $B^* \bar D \to   B^{(*)}\bar D^*$. Consequently, no    elastic OPE-driven interactions  emerge in the low-lying channels  $B \bar D$ for $J=0$ and $B^* \bar D$ for $J=1$.
  On the other hand,    diagonal interactions in these lower channels do receive contributions from two-pion exchange via  intermediate coupled-channel loops, namely $B \bar D \to B^* \bar D^* \to B \bar D$ and $B^* \bar D \to   B^{(*)}\bar D^* \to B^* \bar D$. 
The OPE potentials $12 \to 1'2'$ in the relevant off-diagonal and diagonal $B^{(*)}\bar D^{(*)} \to  B^{(*)}\bar D^{(*)}$  transitions  read  
\bea\nonumber\label{Eq:OPE}
  V_{\text{OPE}}[12 \to 1'2'] &=&  (\bm\tau_1 \cdot \bm\tau_2)\frac{g^2}{4F_{\pi}^2} \; \frac{ ({\bm k\cdot \bm S_1 }  )  ({\bm k \cdot \bm S_2  }) }{ 2\omega_{\pi}({\bm k})} \\ &\times&\left(\frac1{D_{21'\pi }}+\frac1{D_{12'\pi }}\right),
\eea
where $S_i$ $(i=1,2)$ refers either to a polarization vector  $\bm\epsilon_i^{(*)}$  in case of a pseudoscalar-vector-pion vertex or  to  \mbox{$i (\bm\epsilon_i \times \bm\epsilon_{i'}^*)$} for a vector-vector-pion vertex,  $\omega_i = \sqrt{p_i^2 + m_i^2}$,  with \mbox{$\bm p_1=-\bm p_2 \eqqcolon \bm p$} \mbox{($\bm p_{1'}=-\bm p_{2'}\eqqcolon \bm p'$)}
 and $\omega_\pi (k) = \sqrt{k^2 + m_\pi^2}$ with $m_\pi$ being the pion mass and $\bm k=\bm p'+\bm p$. 
 The three-body propagators read 
 \bea
 D_{21'\pi } &=& E -  \omega_2 - \omega_{1'} -\omega_\pi,\\
D_{12'\pi } &=& E -  \omega_1 - \omega_{2'} -\omega_\pi.
 \eea
The isospin factor for the  OPE potential in the isoscalar channels reads $\bm\tau_1 \cdot \bm\tau_2 = -3$.  The dependence of the pion decay constant $F_{\pi}$ on the pion mass is considered using the one-loop chiral perturbation theory following Ref.~\cite{Baru:2013rta},
which gives $F_{\pi}\approx 100$~MeV  for $m_{\pi}=220$ MeV. 
The value of the coupling constant $g$ was extracted from fits to the lattice data of Ref.~\cite{Becirevic:2012pf} and its experimental value at the physical pion mass in \cite{Meng:2023bmz}. For the given lattice spacing of $a \approx 0.12$ fm and $m_{\pi}=220$~MeV, one has $g=0.45$ for $D^{(*)}$ mesons (see Supplemental Material in Ref.~\cite{Meng:2023bmz}), which is consistent with   $g=0.51$  for $B^{(*)}$ mesons~\cite{Flynn:2015xna}. Therefore, in what follows, we use $g=0.45$ for all couplings.

At order $\mathcal{O}(Q^2)$, i.e.~at NLO, momentum-dependent contact interactions start to contribute.  The corresponding Lagrangians can be obtained by acting with 
two derivatives on the Lagrangian in Eq.~\eqref{Lag_LO_CT}, see, e.g.,  Refs.~\cite{Baru:2016iwj,Baru:2019xnh}. 
The higher-order momentum-dependent S-wave interactions can be incorporated through the following replacements in Eqs. \eqref{vfull0+}-\eqref{vfull2+}:
\bea\nonumber
\mathcal{C}_{d} &\to& \mathcal{C}_{d} + \mathcal{D}_{d} (p^2+p'^2)+\cdots ,\\
\mathcal{C}_{f} &\to& \mathcal{C}_{f} + \mathcal{D}_{f} (p^2+p'^2)+\cdots,
\label{eq:CThigh}
\eea
where the dots stand for interactions of order $\mathcal{O}(Q^4)$ and higher.
The contact interactions   and the OPE  are supplemented  with the exponential regulators of the form $e^{\frac{-(p^n+p'^n)}{\Lambda^n}}$  with $n=6$. 

Finally, we emphasize that the energy levels associated with $D$-wave interactions are fully consistent with the noninteracting spectrum. Based on this observation, we restrict our analysis to $S$-wave interactions in what follows. However, in Sec.~\ref{Sec:EFT2}, we also discuss $S$-to-$D$-wave transitions induced by the OPE.

  For the given EFT potential, the FV energy levels (for a given set $J^P$) in a given irreducible representation (irrep) are obtained by solving
the determinant equation  
\begin{equation}
	\det\left[\mathbb{G}^{-1}(E)-\mathbb{V}(E)\right]=0\label{eq:rel_det0},
\end{equation}
which arises from the Lippmann-Schwinger-type integral equations  (LSE) in  FV:
  \be
 \mathbb{T}^{\alpha\beta}(E)=\mathbb{V}^{\alpha\beta}(E)+\mathbb{V}^{\alpha\gamma}(E)\mathbb{G}^{\gamma}(E)\mathbb{T}^{\gamma\beta}(E),
 \ee
where $\alpha, \beta$, and  $\gamma$
label the basis states defined in Eq.~(\ref{Eq:JP}).  
 Here, the potential $\mathbb{V}$ is defined as 
\bea\nonumber
\mathbb{V}^{\alpha\beta}_{\bm{n},\bm{n}'} &=& 4 \sqrt{m_1 m_2 m_{1'}m_{2'}}\; V^{\alpha\beta}(p_{\bm{n}},p_{\bm{n^{\prime}}}), 
\eea
where $V^{\alpha\beta}(p_{\bm{n}},p_{\bm{n^{\prime}}})$ 
represents the sum of the contact and OPE potentials from Eqs.~\eqref{vfull0+}-\eqref{vfull2+},  and \eqref{Eq:OPE}, including the higher-order
terms from Eq.~\eqref{eq:CThigh}. The discretized propagator $\mathbb{G}$ reads
\bea
\mathbb{G^\gamma}_{\bm{n},\bm{n}'}={\cal J}\frac{1}{L^{3}}G^\gamma(p_{\bm{n}},E)\delta_{\bm{n}',\bm{n}},
\eea
 where the Green function $G^\gamma$ is\footnote{The integral equations with this Green function are referred to as the Kadyshevsky equations \cite{Kadyshevsky:1967rs}, see also \cite{Baru:2019ndr} for an alternative derivation and related discussion.}   
\bea\nonumber
G^\gamma(  p,E)=\frac{1}{4\omega_{1,\gamma}(  p^2)\omega_{2,\gamma}(  p^2)}  \frac1{E-\omega_{1,\gamma} (  p^2) -\omega_{2,\gamma} (  p^2) }  ,
\eea
while $\mathcal{J}$ is the Jacobi determinant arising from the transformation between the box and the center-of-mass frames, see \cite{Meng:2023bmz} for details, and $p_{\bm{n}}$ are
the discretized momenta\footnote{Note that $\mathcal{J} = 1$ holds for the lattice irreps $A_1^+[0]$ and  $T_1^+[0]$ from Ref.~\cite{Alexandrou:2023cqg}, since they correspond to the special case of zero total momentum. However, our method can be also applied to moving boxes in the same manner \cite{Meng:2021uhz,Meng:2023bmz}.}. 
Furthermore, the heavy-meson energies
$\omega_{1,\gamma}$ and $\omega_{2,\gamma}$ in the channel $\gamma$ are related with the three-momenta $p$ as 
\be
 \omega_{a,\gamma} (p) = E_{a,\gamma}(0) +  \sqrt{({m_{{a,\gamma}}^{\rm kin}})^2+  p^2}-m_{{a,\gamma}}^{\rm kin},
\ee
where  $a=1,2$ ($a=1$ 
stands for either ${B}$ or ${B^{*}}$ and $a=2$ for $\bar{D}$ or $\bar{D}^{*}$, depending on the channel $\gamma$ in Eq.~\eqref{Eq:JP} for the given $J^P$), and the mass parameters $E_{a,\gamma}(0)$ and $m_{{a,\gamma}}^{\rm kin}$ are taken from Tables II and III in the Supplemental Material of \cite{Alexandrou:2023cqg}.

To solve Eq.~\eqref{eq:rel_det0} in a finite volume, we use the plane wave basis instead of expanding it in partial waves, see~\cite{Meng:2021uhz} for  details.

Once the LECs are adjusted to reproduce the FV spectra (see Secs.~\ref{Sec:EFT1}-\ref{Sec:EFT2} for  details), the IFV amplitudes are obtained by solving  LS-type integral equations. In the single-channel case, the on-shell IFV scattering amplitude can be related to the phase shifts as
\be
\frac{-2\pi}{\mu}T^{-1} (E)= k \cot \delta - i k,  
\ee
with $k$ denoting the on-shell momentum in the center-of-mass frame. 
In the vicinity of the threshold, the phase shifts can be parametrized in terms of the ERE
\be
k \cot \delta = \frac1{a} + \frac12 r k^2 + v_2 k^4 + {\cal O}(k^6),
\ee
where   $a$ and $r$ denote the scattering length and the effective range, respectively, while $v_2$ represents the so-called shape parameter. 
The convergence radius of the ERE is limited by  the nearest left-hand singularity of the scattering amplitude.

In what follows,  we use the naming scheme CT$N_d$+$N_{C_f}$+$N_{D_f}$ for fits involving only contact interactions, where  $N_d$   denotes the number of contact terms included in the diagonal channels, and  $N_{C_f}$ and $N_{D_f}$ denote the number of momentum-independent ${\cal O}(Q^0)$ and momentum-dependent ${\cal O}(Q^2)$ off-diagonal interactions, respectively. When the OPE is included, the corresponding fits are labeled CT$N_d$+$N_{C_f}$+$N_{D_f}${\&}OPE.  
\section{Single channel EFT  at low energies}
\label{Sec:EFT1}

\begin{figure*}[!htbp]
    \centering     
\includegraphics[width=0.4\linewidth]{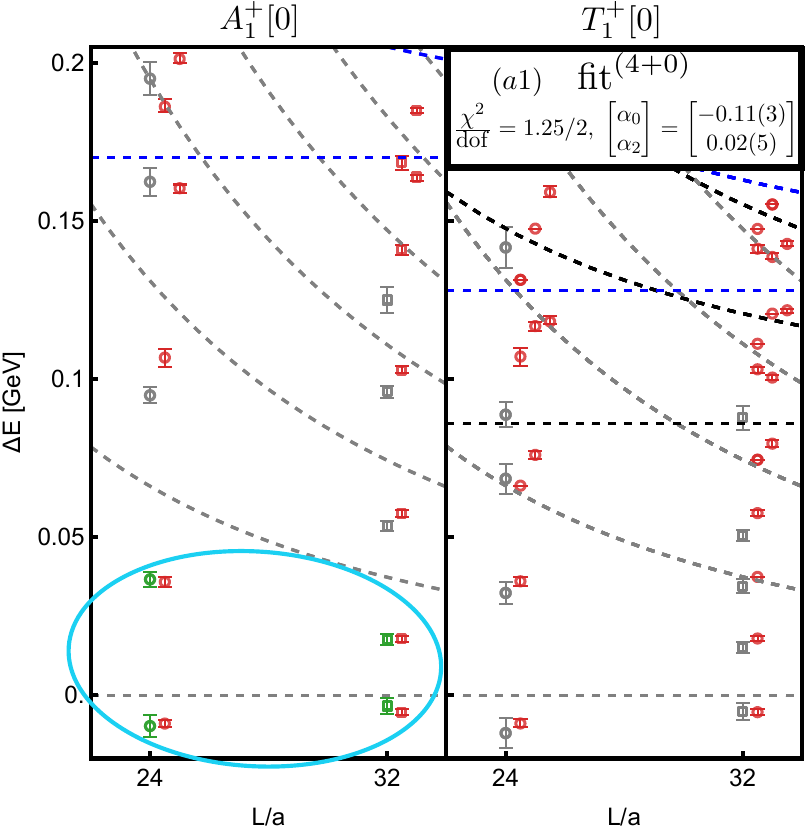}\hspace{0.4cm}
\includegraphics[width=0.55\linewidth]{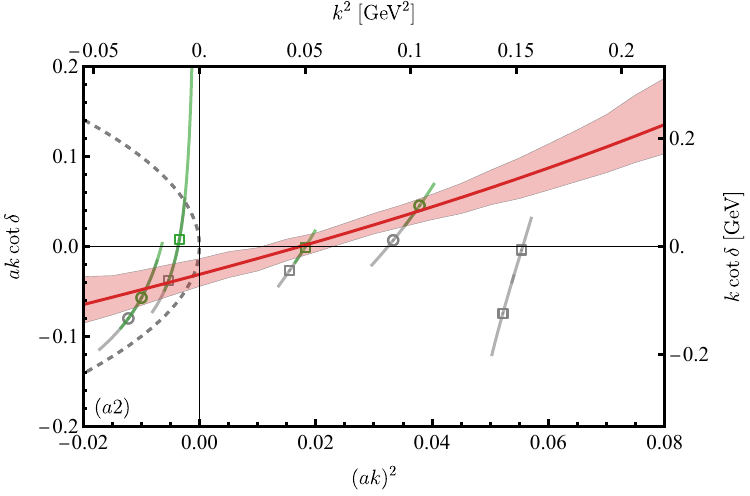}\\ 
\includegraphics[width=0.4\linewidth]{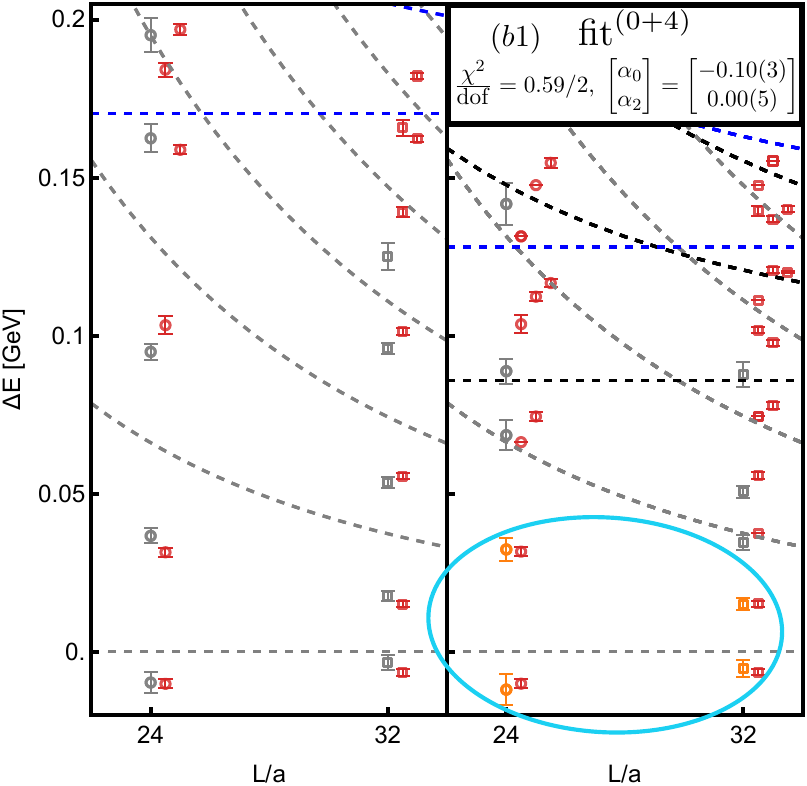}\hspace{0.4cm}
\includegraphics[width=0.55\linewidth]{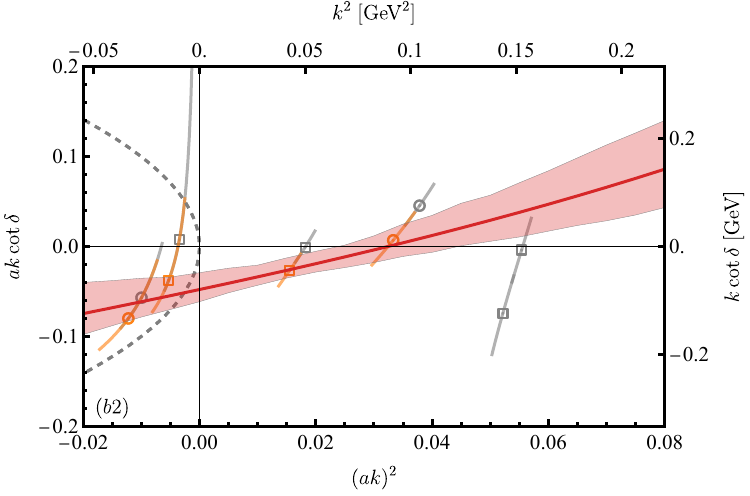}\\ 
\includegraphics[width=0.4\linewidth]{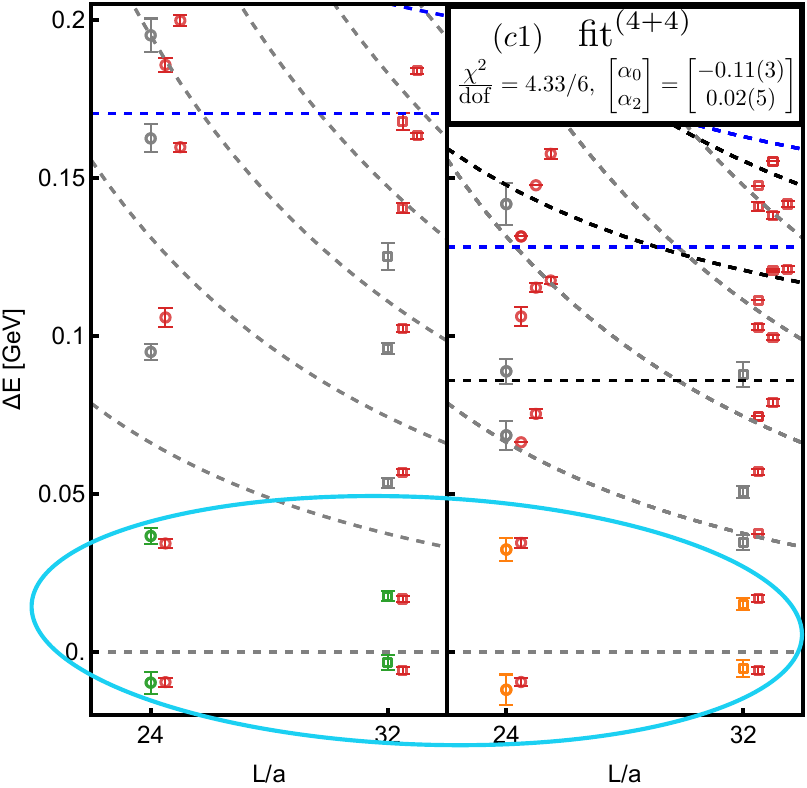}\hspace{0.4cm}
\includegraphics[width=0.55\linewidth]{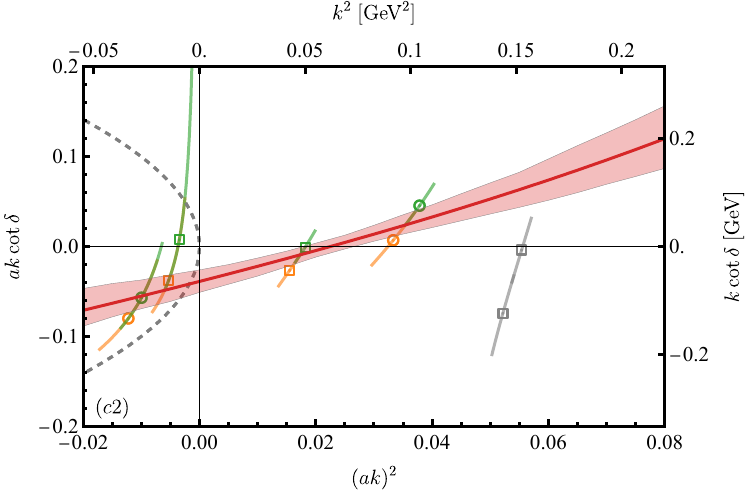}\\
           \caption{ 
           Results of the single-channel EFT (EFT1) for the finite-volume spectra and the infinite-volume S-wave phase shifts $\delta$. 
            Left panel: FV spectra in the $A_1^+[0]$ ($J=0$)  and the $T_1^+[0]$ ($J=1$) irreps. Right panel:    predictions for the infinite-volume $k \cot \delta$. 
           The three fits discussed in the text---fit$^{(4+0)}$, fit$^{(0+4)}$, and fit$^{(4+4)}$---are presented in panels ($a$1), ($b$1), and ($c$1), respectively. Lattice data points from $A_1^+[0]$ used in the fits are shown in green, those from $T_1^+[0]$  in orange, while data points not included in the fits are indicated in gray. 
           The fit range in each fit is also indicated by a cyan ellipse.
           The EFT results are shown in red, with error bands indicating the 68\% ($1\sigma$) statistical uncertainty. Each panel also reports the $\chi^2/{\rm dof}$ and the fitted values of parameters defined in Eq.~\eqref{def_alpha_1}.   
           The noninteracting energy levels (left panel) are shown as dashed lines: gray for the lowest channel ($B\bar D$ or $B^*\bar D$, depending on $J$),
           black for $B\bar D^*$, and blue for $B^*\bar D^*$. 
           The predicted energy levels are displayed up to approximately 30 MeV above the $B^* \bar D^*$ threshold. 
The predicted FV levels from different two-meson channels are horizontally shifted so that levels from each channel are easily distinguishable. For $A_1^+[0]$  irrep, the $B^* \bar D^*$ levels are placed to the right of the $B \bar D$ levels. For $T_1^+[0]$, the ordering from left to right is $B^*\bar  D$, $B\bar D^*$ and $B^*\bar D^*$. 
           The gray    dashed lines in the right panels correspond to $ik = \pm |k|$ from unitarity. }
   \label{Fig_EFT1}
\end{figure*}

\begin{figure}[t]
    \centering    
\vspace*{0.7cm}
\raisebox{0.75cm}
{\includegraphics[width=1.0\linewidth] 
{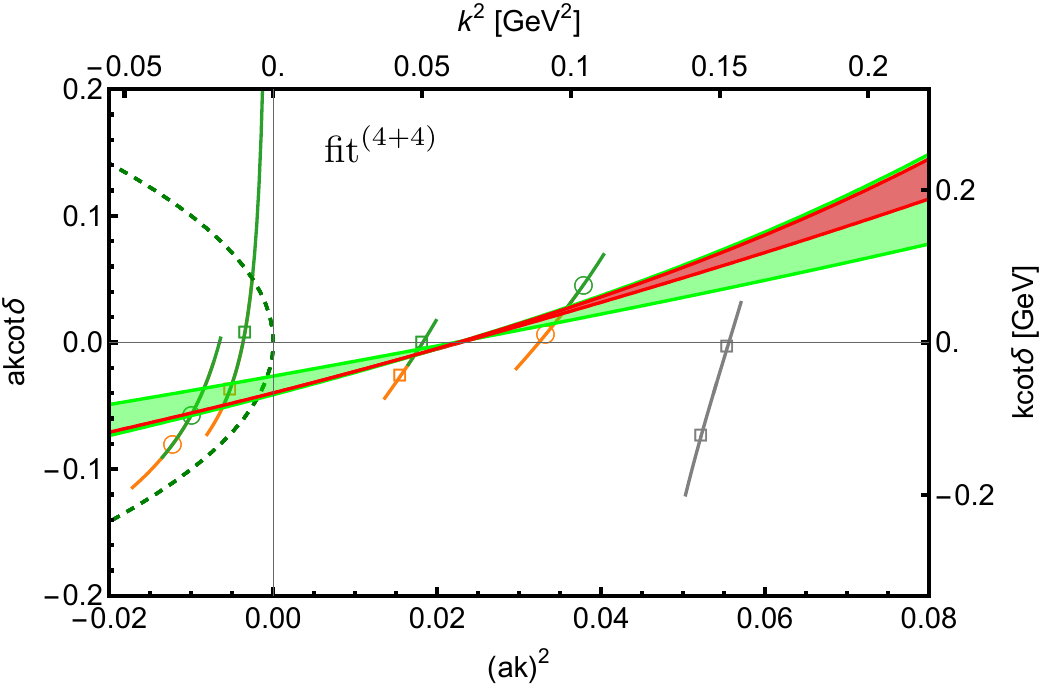}} 
           \caption{Residual cutoff dependence of $k \cot \delta$  from fits with one (CT1+0+0, green band) and two (CT2+0+0, red band) contact terms in EFT1. 
           For $k \cot \delta >0$,  
           the upper edge of each band corresponds to the lowest cutoff $\Lambda= 0.8$~GeV and the lower edge  to the highest cutoff $\Lambda = 1.2$ GeV; for $k \cot \delta < 0$, the ordering is reversed.      The green dashed line corresponds to $ik = \pm |k|$ from unitarity. } 
   \label{Fig:kcotd_cutoff}
\end{figure}

In the first approach, we treat the momentum scales associated with coupled-channel effects as hard and integrate them out. This leads to a low-energy EFT valid near the lowest thresholds, namely the $B \bar D$ threshold for $J=0$ and the $B^* \bar D$ threshold for $J=1$.
As discussed above, the OPE does not contribute directly in this regime, since it is governed by the coupled-channel momentum scales.
As a result, we employ a purely contact, single-channel theory, with LECs adjusted to fit the lowest-lying FV energy levels. Specifically, we consider either three data points (corresponding to the maximum momentum $k_{\text{max}} \simeq 310$~MeV) or four data points ($k_{\text{max}} \simeq 380$ MeV) in each channel.
It is also worth noting that for the considered pion mass, the left-hand cut associated with the $\rho$-meson exchange begins at $k = i m_\rho / 2 \simeq i 400$~MeV. We, therefore, consider the momentum range below this scale.

HQSS dictates that the elastic interactions in the low-lying $B \bar D$ and $B^* \bar D$ channels are identical (see the ``11'' matrix elements in Eqs.~\eqref{vfull0+} and \eqref{vfull1+}):
\be
\label{Eq:CT}
\hspace{-0.1cm}V[B \bar D ] = V[B^* \bar D  ] = \mathcal{C}_{0}+ \mathcal{C}_{2} (p^2+p'^2) +  \mathcal{C}_{4} (p^4+p'^4),
\ee
where, following Eq.~\eqref{eq:CThigh}, we extended the interactions to incorporate   terms of order $\mathcal{O}(Q^2)$ and $\mathcal{O}(Q^4)$, while those of $\mathcal{O}(Q^6)$ and higher  are omitted.
Following \cite{Abolnikov:2024key}, we can rewrite the LECs in the following form,
\bea\label{def_alpha_1}
{\cal C}_0 = \frac{\alpha_0}{F_{\pi}^2}, \hspace{0.3 cm} 
   {\cal C}_2 = \frac{\alpha_2}{F_{\pi}^2}\frac{1}{\Lambda_\chi^2}, \hspace{0.3 cm}   
    {\cal C}_4= \frac{\alpha_4}{F_{\pi}^2} \frac{1}{\Lambda_\chi^4}, 
 \eea
  where $\{\alpha_0, \,\alpha_2, \,\alpha_4\}$ are dimensionless parameters adjusted to obtain the best fit and $\Lambda_{\chi} \simeq 1.0$ GeV is the chiral symmetry breaking scale.

We test HQSS through three independent fits:
(a) a fit to the four lowest data points in $A_1^+[0]$ referred to as fit$^{(4+0)}$,
(b) a fit to the four lowest data points in $T_1^+[0]$ referred to as fit$^{(0+4)}$, and
(c) a combined fit to the four lowest points in both $A_1^+[0]$ and $T_1^+[0]$ referred to as fit$^{(4+4)}$.
All other FV energy levels in $A_1^+[0]$ and $T_1^+[0]$ come out as predictions.

The results of these fits for the FV spectra for $A_1^+[0]$ and  
$T_1^+[0]$ are shown in the left panel of Fig.~\ref{Fig_EFT1}, while the predicted IFV phase shifts ($k \cot   \delta$) are shown in the right panel. All results correspond to fits with two contact terms (CT2+0+0), $\mathcal{C}_0$ and $\mathcal{C}_2$, using the cutoff value of $\Lambda = 1.0$ GeV.

The resulting FV spectra from these fits are nearly identical,  
and the extracted LO contact terms agree within  $\sim 10\%$ (see Fig.~\ref{Fig_EFT1}), indicating only minor possible HQSS violation. 
Table~\ref{Table:EFT1_EFT2_results} summarizes the extracted pole positions in the lowest channels together with the ERE parameters.
In addition to the statistical uncertainty included in all fits, for the HQSS-based fit -- fit$^{(4+4)}$ -- we account for possible HQSS-breaking effects by assigning a 10\% uncertainty to the LO interaction ${\cal C}_0$ and propagating it to the observables.
Furthermore, since the elastic interaction in the $B \bar D^*$ channel is governed by the same linear combination of the contact terms as in the $B \bar D$ and $B^* \bar D$ channels
\be
V[B \bar D^* ] =V[B \bar D ] = V[B^* \bar D  ], 
\label{VBD*}
\ee
as follows from Eqs.~\eqref{vfull0+} and \eqref{vfull1+}, HQSS allows us to predict, without introducing new parameters, the pole position near the $B \bar D^*$ threshold -- see Table~\ref{Table:EFT1_EFT2_results}. 
Given the excellent agreement between the pole positions for the $B \bar D$ and $B^* \bar D$ bound states, a shallow state near the $B \bar D^*$ threshold emerges as a robust prediction that can be verified in future lattice investigations.
 While states near $B^* \bar D^*$ threshold depend on different LEC combinations ($\mathcal{C}_d$, $\mathcal{C}_f$), an approximate but reasonable estimate remains possible under the assumption that the off-diagonal interaction $\mathcal{C}_f$ is
negligible as compared to  $\mathcal{C}_d$. 
The corresponding pole positions are also listed in Table~\ref{Table:EFT1_EFT2_results}. This assumption is supported by the explicit coupled-channel calculation presented in the next section, where we also discuss the corresponding pole structures. 
 
To investigate systematic uncertainties,   Fig.~\ref{Fig:kcotd_cutoff} compares results from fits CT1+0+0 and CT2+0+0 with  one (i.e., only   $\mathcal{C}_0$) and two contact terms ($\mathcal{C}_0$ and $\mathcal{C}_2$), respectively, showing also their variation with the cutoff $\Lambda$ in the range 0.8–1.2 GeV.
For the fit CT1+0+0, the effective range is positive and scales inversely with the cutoff, resulting in some cutoff dependence of $k  \cot \delta$ near threshold.
In contrast, the fit CT2+0+0 removes this dependence through the cutoff running of $\mathcal{C}_2$, yielding nearly cutoff-independent results for the scattering length, the effective range and thus also for the pole position -- see Fig.~\ref{Fig:kcotd_cutoff}.
A mild cutoff dependence beyond the relevant momentum range reflects the influence of higher-order terms.
Since $k \cot \delta$ exhibits essentially linear behavior in $k^2$, the inclusion of the higher-order term $\sim \mathcal{C}_4$ (fit CT3+0+0) does 
not improve the fit quality and instead leads to overfitting. 
Moreover, due to this linear behavior, the choice between using the three or four lowest data points as input has only a minor impact on the results. In particular, the effective range extracted from the fit$^{(3+3)}$ is slightly smaller but remains consistent within statistical errors with that from the fit$^{(4+4)}$.
 
\begin{table*}[t]
\begin{tabular*}{\hsize}{@{}@{\extracolsep{\fill}}lccccccc@{}}
\hline\hline
& Fit Type &  
 Threshold & $J^{P}$ & { Poles {[}MeV{]}} & $1/a$ {[}fm$^{-1}${]} &  
 $r$ {[}fm{]} & $v_{2}$ {[}fm$^{3}${]}\tabularnewline
\hline 
\multirow{6}{*}{EFT1 (CT2+0+0)} & fit$^{(4+0)}$ & $B\bar{D}$ & $0^{+}$ & $-1.1_{-1.6}^{+0.9}$ &  
$-0.27_{-0.14}^{+0.14}$
& $0.41_{-0.15}^{+0.15}$ &  
-----\tabularnewline
\cline{2-8} \cline{3-8} \cline{4-8} \cline{5-8} \cline{6-8} \cline{7-8} \cline{8-8} 
 & fit$^{(0+4)}$ & $B^{*}\bar{D}$ & $1^{+}$ & $-2.6_{-3.0}^{+1.9}$ & 
 $-0.40_{-0.17}^{+0.17}$ &
$0.33_{-0.25}^{+0.25}$ & 
-----\tabularnewline
\cline{2-8} \cline{3-8} \cline{4-8} \cline{5-8} \cline{6-8} \cline{7-8} \cline{8-8} 
 & \multirow{4}{*}{fit$^{(4+4)}$} & $B\bar{D}$ & $0^{+}$ & $-1.8_{-1.3-4.0}^{+1.0+1.6}$ & $-0.33_{-0.10-0.25}^{+0.10+0.24}$ & $0.39_{-0.11-0.02}^{+0.11+0.02}$ & 
 -----\tabularnewline
 &  & $B^{*}\bar{D}$ & $1^{+}$ & $-1.7_{-1.3-4.0}^{+0.9+1.6}$ & $-0.33_{-0.10-0.25}^{+0.10+0.24}$ & $0.39_{-0.11-0.02}^{+0.11+0.02}$ & 
 -----\tabularnewline
 &  & $B\bar{D}^{*}$ & $1^{+}$ & $-3.6_{-1.5-5.0}^{+1.2+2.7}$ & $-0.48_{-0.07-0.23}^{+0.07+0.22}$ & $0.39_{-0.10-0.01}^{+0.10+0.01}$ & 
 -----\tabularnewline
 &  & ${B^{*}\bar{D}^{*}}^{\dagger}$ & $0^{+},1^{+},2^{+}$ & $-3.6_{-1.5-5.0}^{+1.2+2.7}$ & $-0.47_{-0.07-0.23}^{+0.07+0.22}$ & $0.39_{-0.10-0.01}^{+0.10+0.01}$ & 
 -----\tabularnewline
\hline 
\multirow{4}{*}{EFT2 (CT3+0+0)} & \multirow{4}{*}{fit$^{(9+5)}$} & $B\bar{D}$ & $0^{+}$ & $-0.8_{-1.1-0.6}^{+0.6+0.4}$ & $-0.23_{-0.12-0.07}^{+0.12+0.07}$ & $0.37_{-0.11-0.02}^{+0.11+0.02}$ & $-0.025_{-0.007-0.001}^{+0.007+0.001}$\tabularnewline
 &  & $B^{*}\bar{D}$ & $1^{+}$ & $-0.7_{-1.0-0.6}^{+0.6+0.4}$ & $-0.22_{-0.12-0.07}^{+0.12+0.07}$ & $0.38_{-0.11-0.02}^{+0.11+0.02}$ & $-0.025_{-0.007-0.001}^{+0.007+0.001}$\tabularnewline
 &  & $B\bar{D}^{*}$ & $1^{+}$ & $-8.0_{-3.4-1.4}^{+2.8+1.3}$ & $-0.70_{-0.11-0.05}^{+0.11+0.05}$ & $0.25_{-0.08-0.01}^{+0.08+0.01}$ & $-0.024_{-0.006-0.001}^{+0.006+0.001}$\tabularnewline
 &  & $B^{*}\bar{D}^{*}$ & $0^{+},1^{+},2^{+}$ & $-7.6_{-3.3-1.4}^{+2.7+1.3}$ & $-0.68_{-0.11-0.05}^{+0.11+0.06}$ & $0.26_{-0.08-0.01}^{+0.08+0.01}$ & $-0.024_{-0.006-0.001}^{+0.006+0.001}$\tabularnewline
 \hline\hline
\end{tabular*}\caption{Pole positions of the near-threshold $B^{(*)}\bar{D}^{(*)}$ states and  the ERE parameters in the corresponding channels
extracted from the lattice data, together with the HQSS-based predictions. All poles correspond to bound states. 
In the EFT1 fit$^{(4+4)}$, the $B\bar{D}^{*}$ $(1^{+})$ state is
predicted parameter-free from HQSS, while the $B^{*}\bar{D}^{*}$
states, marked with $\dagger$, are obtained under the additional
assumption that off-diagonal interactions are negligible, as supported
by the EFT2 results. 
The first uncertainty represents the statistical $1\sigma$ error, while the second (for fits (4+4) and (9+5)) accounts for a 10\% HQSS-violation in the LO interaction ${\cal C}_0$ in Eq.~\eqref{Eq:CT}. 
The EFT1 results are shown for a cutoff of $\Lambda=1.0$ GeV, whereas for EFT2 we use $\Lambda=1.2$ GeV, which ensures a clearer separation of scales. We note that the cutoff-induced uncertainties in the pole positions are much smaller than the other uncertainties quoted in the Table -- see Fig.~\ref{Fig:kcotd_cutoff}
and Fig.~\ref{fig:3CT_cutoff} in Appendix \ref{App:EFT2_results} for further details. 
We do not show values for $v_2$ in the EFT1 case, as they are consistent with zero and cannot be reliably extracted from low-energy fits.
}
\label{Table:EFT1_EFT2_results}  
\end{table*}

\section{EFT in the whole momentum range}
\label{Sec:EFT2}

\subsection{Coupled-channel calculations of the finite volume energy levels}
\label{Sec:coupled_EFV}

\begin{figure*}[t]
    \centering    
{\includegraphics[width=0.45\textwidth]{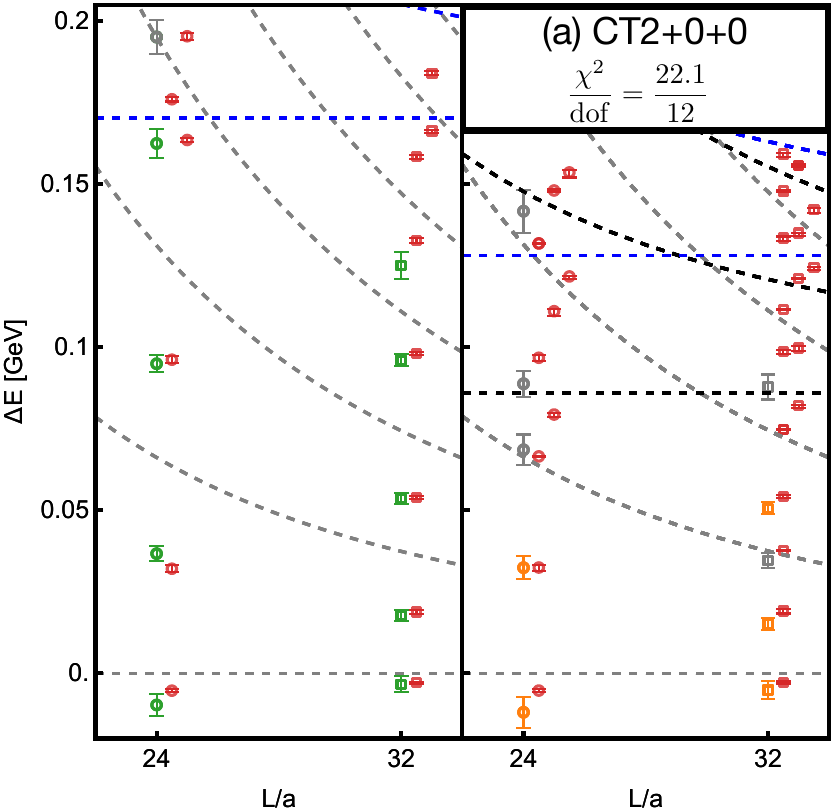}} 
{\includegraphics[width=0.45\textwidth]{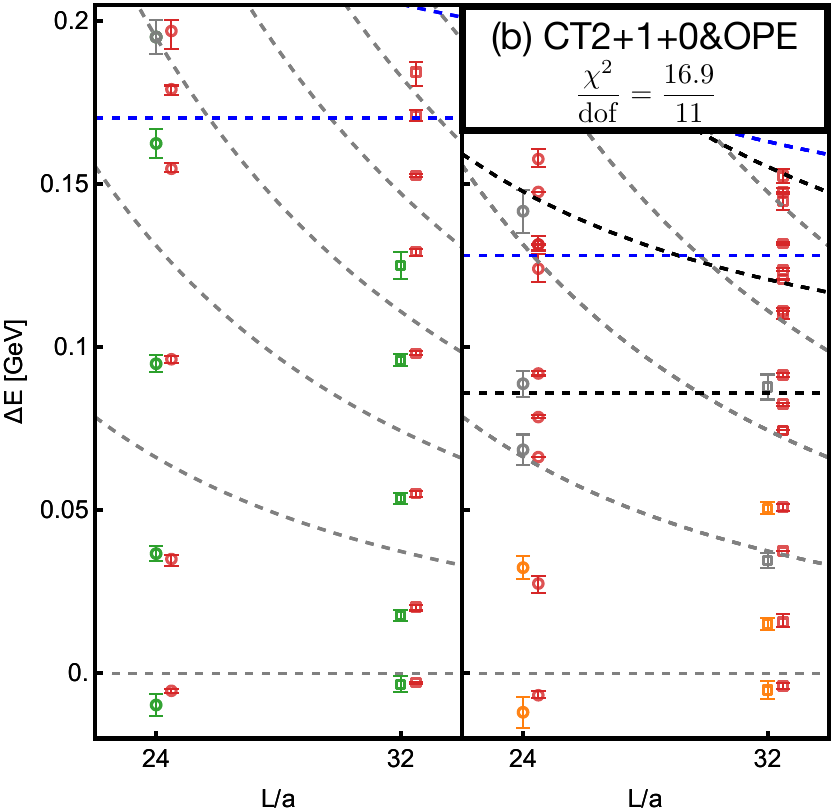}} \\
{\includegraphics[width=0.45\textwidth]{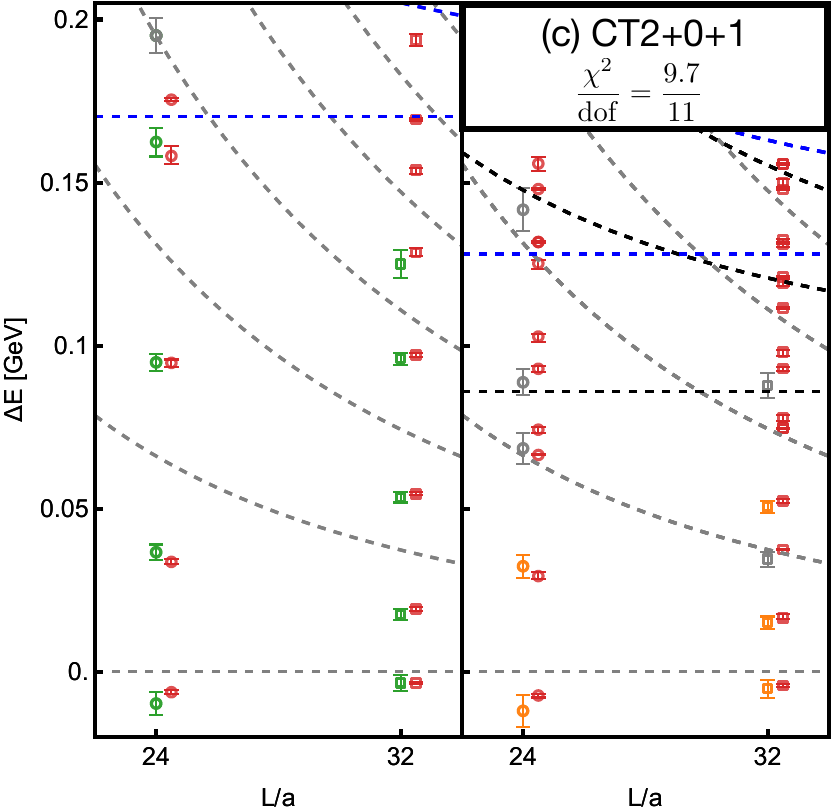}}  
{\includegraphics[width=0.45\textwidth]
{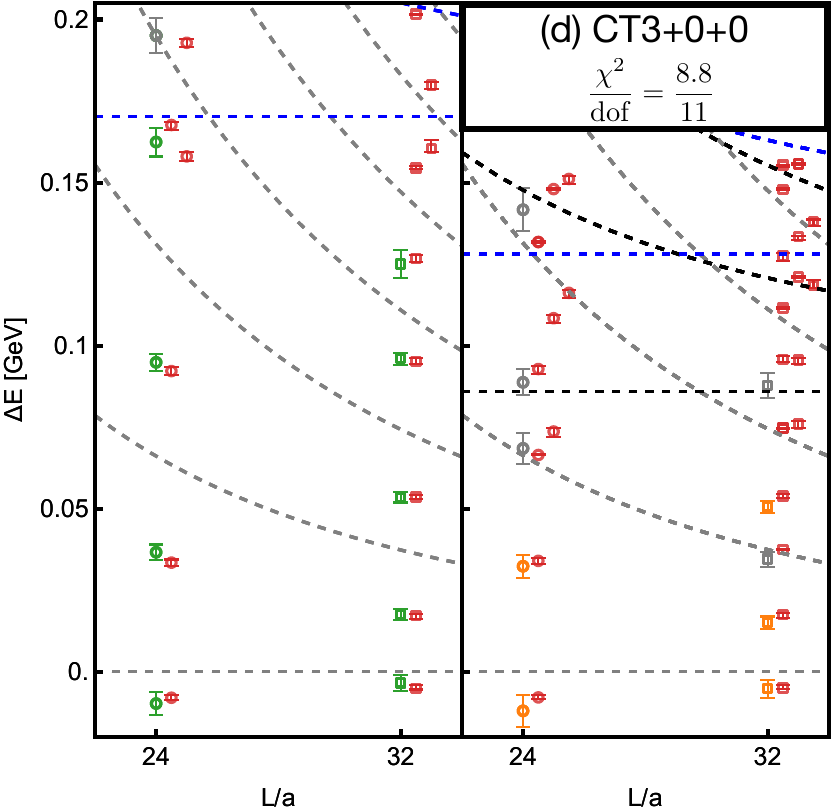}}  \\
           \caption{Comparison of the EFT2 results obtained using different interaction terms.
           Panel (a) shows the fit with two diagonal contact terms (CT2+0+0); panel (b) displays the fit including the OPE interaction together with the off-diagonal term  ${\cal C}_f$ (CT2+1+0{\&}OPE); panel (c) shows the fit including the off-diagonal term ${\cal D}_f$ (CT2+0+1) and panel (d) presents the fit with three diagonal contact terms (CT3+0+0), as discussed in text. All results correspond to a cutoff of 1.2 GeV. The color coding is the same as in Fig.~\ref{Fig_EFT1}.  For fits with diagonal interactions, as in Fig.~\ref{Fig_EFT1}, FV levels from different two-meson channels are horizontally shifted so that levels from each channel are easily distinguishable. See the caption of Fig.~\ref{Fig_EFT1} for further details.
         }
   \label{Fig:EFT2_EFVcc}
\end{figure*}

 In the second approach, the momentum scales associated with coupled-channel effects are treated as soft. Accordingly, we allow for coupled-channel interactions driven by both one-pion exchange  and contact terms. This framework is, to our knowledge, the first coupled-channel EFT-based analysis of FV spectra. It enables us to  incorporate  the OPE dynamics and the associated left-hand cuts  into the analysis of the lowest-lying channels. Assuming that the effects of $\rho$-meson exchange are largely represented by 
contact interactions, we include all available lattice data points between the lowest and higher thresholds in the fits.

The coupled-channel calculations are carried out step by step as follows:

1. We begin with a 
contact EFT including two diagonal contact interactions (fit CT2+0+0), fitted simultaneously to both the 
$A_1^+[0]$ and  $T_1^+[0]$ irreps using all relevant data points up to the next threshold
(i.e., 9 points in $A_1^+[0]$ and 5 in $T_1^+[0]$).
The highest-energy data point in this fit (the 9th point in $A_1^+[0]$) corresponds to the shallow ground-state level associated with the $B^* \bar D^*$ threshold.
Excluding this point from the fit yields virtually the same result, with this ground-state level emerging as a prediction.

While this setup generally provides a reasonable description of the FV spectrum (see Fig.~\ref{Fig:EFT2_EFVcc}(a)), it fails to quantitatively reproduce the data near the lowest thresholds. Furthermore, the QQ plots presented in Appendix~\ref{QQapp} reveal significant deviations from the expected normal distribution of the residuals, thus indicating that such fits may not be regarded as statistically consistent.

2. We then add various coupled-channel interactions involving   both short-range  terms and the long-range OPE and  discuss their impact on the results.
 
 2.1  The CT2+1+0 fit with three contact terms including ${\cal C}_f$  provides only a minor and statistically insignificant improvement   in the $\chi^2$ compared to the  CT2+0+0 fit, 
 while  the QQ plots still demonstrate a pronounced deviation from the normal distribution.

 2.2 The effect of the one-pion exchange   is illustrated in Fig.~\ref{Fig:EFT2_EFVcc}(b) -- fit CT2+1+0{\&}OPE. For the lower channels, $B \bar D$ and $B^* \bar D$, the OPE enters only through off-diagonal terms and contributes solely through the effective elastic  TPE transitions. We note that to properly renormalize the OPE, it must appear together with contact interactions~\cite{Baru:2015nea}. For S-wave transitions, it is the off-diagonal momentum-independent contact term
${\cal C}_f$ that ensures the renormalization.  
For example, for the static OPE, one finds that the linear combination
$\tilde{C}_f = {\cal C}_f + \frac{g^2}{4 F_\pi^2}$ absorbs the leading short-range part of the OPE  into the redefined contact  term. This linear combination then contributes to all potentials with different angular momenta $J$, as required by HQSS.
For dynamical pions, the renormalization is performed  numerically.
Although the inclusion of S-wave pions leads to a further decrease in $\chi^2$,   
the resulting effect is minor as compared to an analogous fit without pions. 
On the other hand, the influence of the OPE is expected to be more noticeable near the  $B^* \bar D^*$ threshold, where it enters as a diagonal repulsive interaction. 
 
To estimate the impact of the D-waves in the OPE on the results, we proceed as follows. Since the role of the  D-waves increases with momentum, we focus on the irrep  $A_1^+[0]$, where energy levels are available over a larger energy range up to the $ B^* \bar D^*$  threshold. We then extend the basis of the $0^+$ states to allow for the D-wave OPE in the $B^* \bar D^*$ channel (note that $B \bar D$ can only occur in the S-wave). Specifically, we incorporate the S-wave-to-D-wave OPE transitions $B\bar D (^1 S_0)\to B^*\bar D^* (^5 D_0)$ and $B^*\bar D^* (^1 S_0)\to B^*\bar D^* (^5 D_0)$ into the current fit CT2+1+0{\&}OPE, and refit the contact interactions.
The resulting effect on the $\chi^2$ and on the energy levels is found to be almost negligible, consistent with the nearly noninteracting D-wave energy levels.

\begin{figure*}
    \centering
     \includegraphics[width=0.4\linewidth]{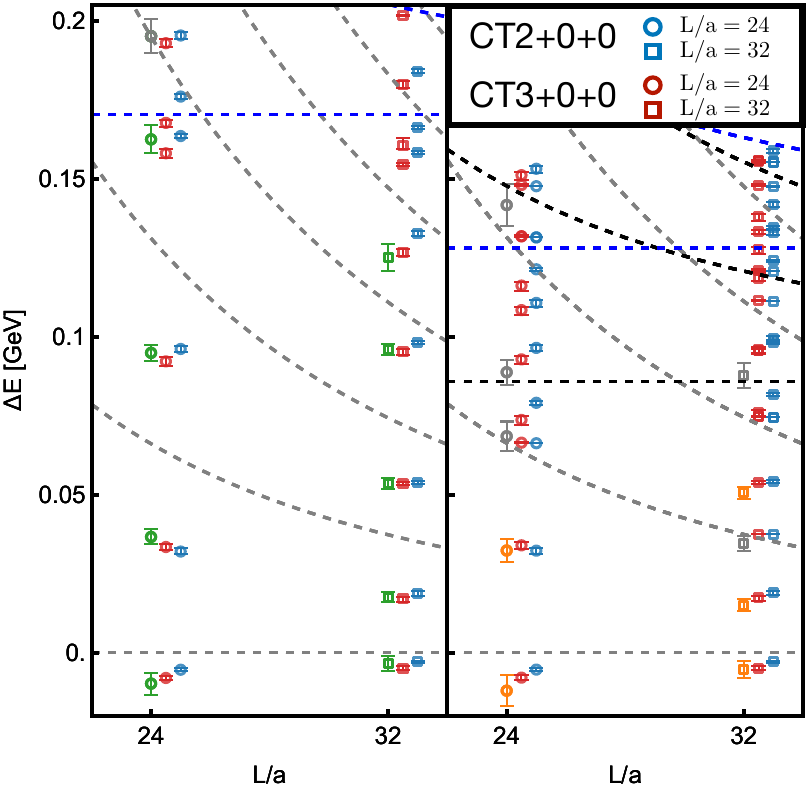}
    \includegraphics[width=0.55\linewidth]{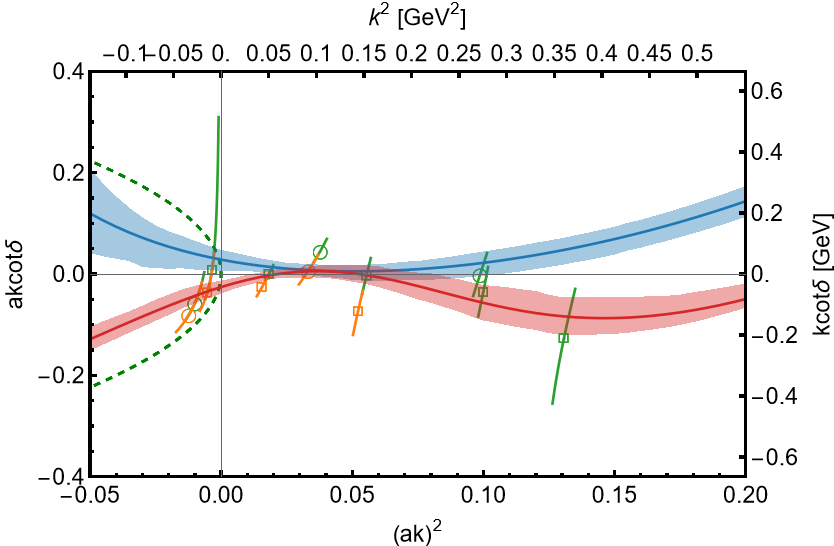}\\
    \caption{ Convergence of the results from fits using  
    two (blue)  
    and three (red)  
    diagonal contact interactions.   The FV spectra in the $A_1^+(0)$ and $T_1^+(0)$  irreps are shown in the left panel. Right panel  shows corresponding   $k \cot \delta$ in the infinite volume.  Bands indicate 1$\sigma$ statistical errors.    
   The results are shown for a cutoff of 1.2 GeV. Lattice data points from $A_1^+[0]$ used in the fits are shown in green, those from $T_1^+[0]$  in orange, while data points not included in the fits are indicated in gray. 
 }
    \label{fig:EFT2_2CT_3CT}
\end{figure*}

 2.3 Fitting the lattice data with four contact terms, including both ${\cal C}_f$ and ${\cal D}_f$ in addition to the two diagonal contact interactions, leads to the appearance of multiple minima close to each other, with the contact terms exhibiting strong correlations among themselves. This behavior points toward overfitting.

We therefore consider an alternative  contact scenario  in which only ${\cal D}_f$ is included in addition to the diagonal contact terms (fit CT2+0+1), as illustrated in   Fig.~\ref{Fig:EFT2_EFVcc}(c). While the contact terms are still  strongly correlated, with the off-diagonal interaction being the largest among all contact terms, the fit exhibits a well separated global minimum with a significantly reduced $\chi^2$.  On the other hand, although the overall description of the data is visibly improved,  the ground-state energy levels at smaller volumes are still not reproduced  with sufficient accuracy.  Furthermore, the QQ plot for this fit still shows significant deviations from the expected normal distribution of the residuals.

3. As discussed above, while the fit CT2+0+0  
provides an overall reasonable description of the data, it fails to reproduce the near-threshold energy levels with high accuracy. This deficiency can not be cured sufficiently well by including off-diagonal interactions.   
Furthermore, in this fit, multiple points in the QQ plot  
lie outside the 68\% interval and some even outside the 95\% interval, indicating that the residuals do not follow a normal distribution.
There is one more subtle issue with this fit:  although the results of the fit CT2+0+0 near the lowest thresholds are stable for cutoffs $\Lambda \gtrsim 1.2$ GeV\footnote{Note that a  cutoff of 1.2 GeV with our regulator corresponds to a sharp cutoff of less than 1.0 GeV.}, the cutoff dependence becomes more pronounced for $\Lambda \lesssim 1$ GeV. 
This behavior likely reflects a poorer separation between soft and hard scales at lower cutoffs, which leads to unphysical poles in $k \cot \delta$ appearing close to threshold and distorting the low-energy behavior. 

 To address these issues, we perform a  three contact term  
 fit by supplementing the diagonal interaction with an additional ${\cal O}(Q^4)$ contact term of the form ${\cal C}_4 (p^4 + p'^4)$  (fit CT3+0+0), see Eq.~\eqref{Eq:CT}, which provides additional flexibility to account for the nontrivial momentum dependence exhibited by the lattice data, starting from the fifth-lowest data point in each irrep, respectively. The results of this fit are shown in  Fig.~\ref{Fig:EFT2_EFVcc}(d). Furthermore, Fig.~\ref{fig:EFT2_2CT_3CT}  illustrates the convergence behavior of the FV spectra and the  phase shift in the infinite volume
 as   two or three contact terms are included in the diagonal potentials (fits CT2+0+0 vs. CT3+0+0). As expected, introducing the additional contact term improves the fit quality, particularly near the lower thresholds, while also significantly reducing the cutoff dependence -- 
 see Fig.~\ref{fig:3CT_cutoff} in Appendix \ref{App:EFT2_results}, where the results of  fit CT3+0+0 are shown for different cutoffs. 
  As a result,  the cutoff-induced uncertainties in the extracted pole positions are much smaller than other uncertainties quoted in  Table~\ref{Table:EFT1_EFT2_results}.
Still, at higher momenta $(ak)^2 \gtrsim 0.1$, the influence of nearby poles in $k \cot \delta$ 
becomes more substantial, which leads to an increased systematic uncertainty stemming from the cutoff variation.

Although the fits CT2+0+1 and CT3+0+0 in Fig.~\ref{Fig:EFT2_EFVcc} yield very similar values of  $\chi^2$, the latter (diagonal) fit  provides a somewhat better description of the ground-state energy levels. In particular, the corresponding $\chi^2$ values for the four ground-state data points are $2.11$ and $1.47$ for these two fits, respectively. 
The resulting infinite-volume ground-state poles are very shallow virtual states in fit CT2+0+1, but correspond to bound states in fit CT3+0+0. 
In addition, distribution of residuals of the fit  CT3+0+0 is consistent with a normal one, in contrast to the fit  CT2+0+1 --- see the QQ plots in Fig.~\ref{Fig:QQall}  
in Appendix~\ref{QQapp}.   Although the available lattice data do not permit an unambiguous discrimination between these two fits, they result in notably different predictions for the energy levels above the $B\bar D^*$ threshold, most prominently near the $B^*\bar D^*$ threshold in the $T_1^+[0]$ ($J=1$) irrep for $L=24$.
Future lattice data in the higher channels would therefore provide valuable additional constraints on the effective interactions.

To further assess the stability of our results and estimate systematic uncertainties, we  performed several consistency checks:

(a) We  performed a fit CT3+0+0 using an alternative ${\cal O}(Q^4)$ contact term of the form ${\cal C}_4' p^2 p'^2$. Although this term is on-shell equivalent to ${\cal C}_4 (p^4 + p'^4)$, the two exhibit different off-shell behavior, which can introduce systematic effects in observables that can be used to estimate the impact of neglected higher-order terms~\cite{Reinert:2017usi}. 
We found that the resulting $k \cot \delta$ is consistent between both interaction forms within the uncertainties already estimated from the cutoff variation.  
(b)  We performed a self-consistency check by fitting only the $J=0$ data in $A_1^+[0]$ and predicting the $J=1$ spectrum in $T_1^+[0]$, and vice versa. Since HQSS is expected to hold well in this system (as demonstrated in EFT 1), the results should not depend on the choice of fitting input. This consistency is indeed observed for the   fits 
CT2+0+1 and CT3+0+0, as well as for CT2+0+0 -- see, e.g.,  Fig.~\ref{fig95_vs_90} in Appendix~\ref{App:EFT2_results} for the results of the main fit CT3+0+0.

(c) Finally, we have verified that extending the fit CT3+0+0 by including the OPE together with an additional contact term
${\cal C}_f$  (fit CT3+1+0{\&}OPE) leads to only a very minor, perturbative change in the parameter values and, consequently, to an insignificant improvement in the fit quality.  

To summarize, we take the fit CT3+0+0 as our main result among the fits to all lattice data points over the entire momentum range.

\subsection{Infinite-volume poles in the $B^{(*)} \bar D^{(*)}$ channels}  
\label{Sec:EFT2_poles}

In Table \ref{Table:EFT1_EFT2_results}, we present the pole positions in the relevant $B \bar D$ and $B^* \bar D$ channels extracted from the fit CT3+0+0, together with the ERE parameters. In addition, we provide predictions for the pole positions of their heavy-quark spin symmetry partners near the $B \bar D^*$ and $B^* \bar D^*$ thresholds and the corresponding ERE parameters, including the estimated uncertainties. 

The pole positions obtained using both EFT1 and EFT2 are found to be mutually consistent, providing strong support to the existence of $T_{bc}$ bound states identified in Ref.~\cite{Alexandrou:2023cqg} and    yielding robust predictions for their HQSS partners. 
To gain further insight into the nature of the extracted $T_{bc}$ poles, we estimate their compositeness, $\bar{X}_A$ (with $\bar{X}_A = 1$ corresponding to a pure molecular state and $\bar{X}_A = 0$ to a compact configuration), as introduced in Ref.~\cite{Matuschek:2020gqe} to characterize near-threshold bound states, virtual states, and resonances. For all poles listed in Table~\ref{Table:EFT1_EFT2_results}, we find $\bar{X}_A \approx 0.9$, which strongly supports their molecular interpretation.

To account for the peculiar momentum dependence of the lattice data beyond the threshold region considered in EFT1, a strong interplay between attractive and repulsive interactions is required. This interplay leads to the appearance of poles in $k \cot \delta$, corresponding to zero crossings in the phase shifts (see, e.g., the rapidly increasing $k \cot \delta$ at larger momenta for cutoff values of $1.0$~GeV and, especially, $0.8$~GeV in Fig.~\ref{fig:3CT_cutoff}  in Appendix \ref{App:EFT2_results}).
 In addition, resonance poles in the scattering amplitude appear deeply in the complex plane, as also noted in Ref.~\cite{Alexandrou:2023cqg}; however, their precise positions are highly sensitive to details of the EFT setup—such as the number of contact terms and the choice of cutoff---and therefore cannot be regarded as a reliable physical prediction.

\section{Summary and Conclusions}
\label{Sec:summary}

We develop an EFT framework to study the dynamics of coupled-channel $ B^{(*)} \bar D^{(*)}$ scattering, and apply it to recent lattice QCD results from Alexandrou et al. \cite{Alexandrou:2023cqg}. Our analysis is based on two complementary EFT formulations: 
(1) A low-energy theory applicable near the $B \bar D$ ($J=0$) and $B^* \bar D$ ($J=1$) thresholds, where coupled-channel effects are integrated out and encoded in diagonal short-range interactions (EFT1);
(2) A fully coupled-channel formulation (EFT2), in which all relevant momentum scales are treated as soft, incorporating both contact interactions and the one-pion exchange (OPE). In this framework, the OPE contributes to the $B \bar D$   and $B^* \bar D$ channels only through off-diagonal transitions, generating a left-hand cut associated with the two-pion exchange. EFT2 is aimed at exploring the behavior of the lattice spectrum over an extended energy range.

The lowest-lying lattice FV energy levels are internally consistent and accurately described within EFT1 using two HQSS-constrained contact interactions. Independent fits to the  lowest four data points in either the $J=0$ (fit 4+0) or $J=1$ (fit 0+4) channel, as well as a combined fit to both (fit 4+4), yield consistent low-energy constants and predict similar bound-state poles. From these fits, possible violations of HQSS are estimated to be at most $\sim 10\%$, and this is propagated to estimate a systematic uncertainty in the extracted pole positions. 
The extracted low-energy parameters, scattering lengths and effective ranges,  are found to be essentially independent of the cutoff. 
Nearly identical results (within statistical errors) are obtained when only the lowest three data points are used.

Above the fourth data point, however,  the lattice data exhibit significant deviations from the simple near-threshold behavior: while the lowest energy levels follow an approximately linear rise with $k^2$ of $k \cot \delta$, the fifth and higher levels bend downwards, showing a clear departure from linearity and indicating the onset of additional dynamics.
The minimal setup with two diagonal contact terms, as implemented in EFT1, yields a globally reasonable description of all lattice data; however, it does not capture the near-threshold behavior with sufficient accuracy, exhibits increased sensitivity to the cutoff and leads to a distribution of residuals that deviates significantly from a normal distribution.
Attempts to improve the description by including off-diagonal interactions---either contact or OPE-driven  as implemented in EFT2---typically yield multiple nearby minima 
without significantly improving the fit quality. To better capture the observed trend, we  revert to a diagonal-only interaction scheme and introduce a higher-order $\mathcal{O}(Q^4)$ contact term, which provides additional flexibility to account for the nontrivial momentum dependence exhibited by the lattice data. The resulting three-parameter fit leads to the improved description of FV spectra and $k \cot \delta$, particularly near the lowest thresholds, while significantly reducing the  cutoff dependence. This fit is also fully statistically consistent.  Nevertheless, reproducing the full lattice spectrum requires additional fine-tuning of low-energy parameters, making such fits more sensitive to systematic uncertainties than those restricted to the low-energy regime.

The pole positions extracted from both EFT1 and EFT2 remain consistent, reinforcing the existence of shallow $T_{bc}$ bound states reported in Ref.~\cite{Alexandrou:2023cqg} and providing reliable predictions for their HQSS partners. In particular, HQSS  requires the presence of a shallow bound state near the $B \bar D^*$ threshold.  Additional shallow bound states are also expected near the $B^* \bar D^*$ threshold with $J^P=0^+, 1^+$ and $2^+$. These predictions (i) follow directly from the well-constrained low-energy fits, (ii) are nearly cutoff independent, and (iii) do not involve any additional free parameters. They  are amenable to direct tests in future lattice simulations or experiments.

The nontrivial momentum dependence observed beyond the fourth data point in each channel suggests a subtle interplay between attractive and repulsive forces, leading to zero crossings in the phase shift (poles in $k \cot \delta$). While resonance poles do emerge --- though located deeply in the complex plane, as noted in Ref.~\cite{Alexandrou:2023cqg}--- their positions are highly sensitive to model details and cannot be extracted reliably.

Additional lattice data at different pion masses would be crucial to constrain the pion-mass dependence of short-range interactions, which is necessary for controlled chiral extrapolations to the physical point. Moreover, data near the higher thresholds would provide an important test of our EFT predictions and further constrain  short-range interactions. 

\medskip
 
 \begin{acknowledgments}
 We thank Stefan Meinel for providing us with the results of the lattice simulations and useful comments. 
This work was supported in part by the European Research Council (ERC) under the European Union’s Horizon 2020 research and innovation programme (grant agreement No. 885150), by MKW NRW under the funding code NW21-024-A, and by Start-up Funds of Southeast University (Grant No. 4007022506).
\end{acknowledgments}

\bibliography{Tbc_refs}

\appendix

\section{Testing Normality of Fit Residuals with Quantile-Quantile Plots} 
\label{QQapp}

\begin{figure}[htbp]
    \centering    
{\includegraphics[width=0.9\linewidth]
{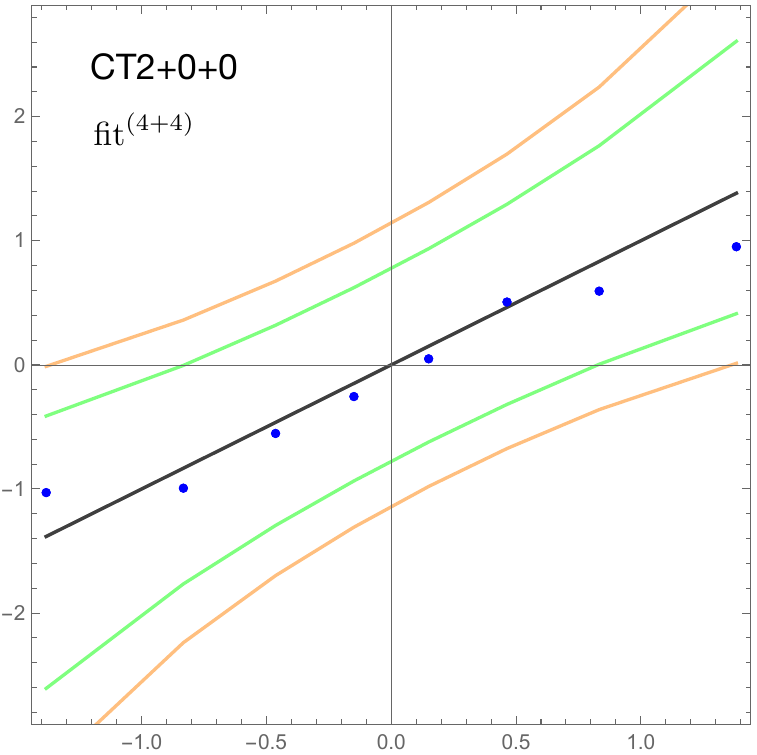}} 
           \caption{ \label{Fig:QQ4+4} QQ plot  for the fit$^{(4+4)}$  EFT1 calculation (blue dots), as described in the main text.  A straight  45$^\circ$ line is shown to guide the eye. The area between the green (orange) lines corresponds to the 68\% (95\%) confidence interval. 
           }  
\end{figure}

\begin{figure*}[htbp]
    \centering    
{\includegraphics[width=0.9\linewidth]{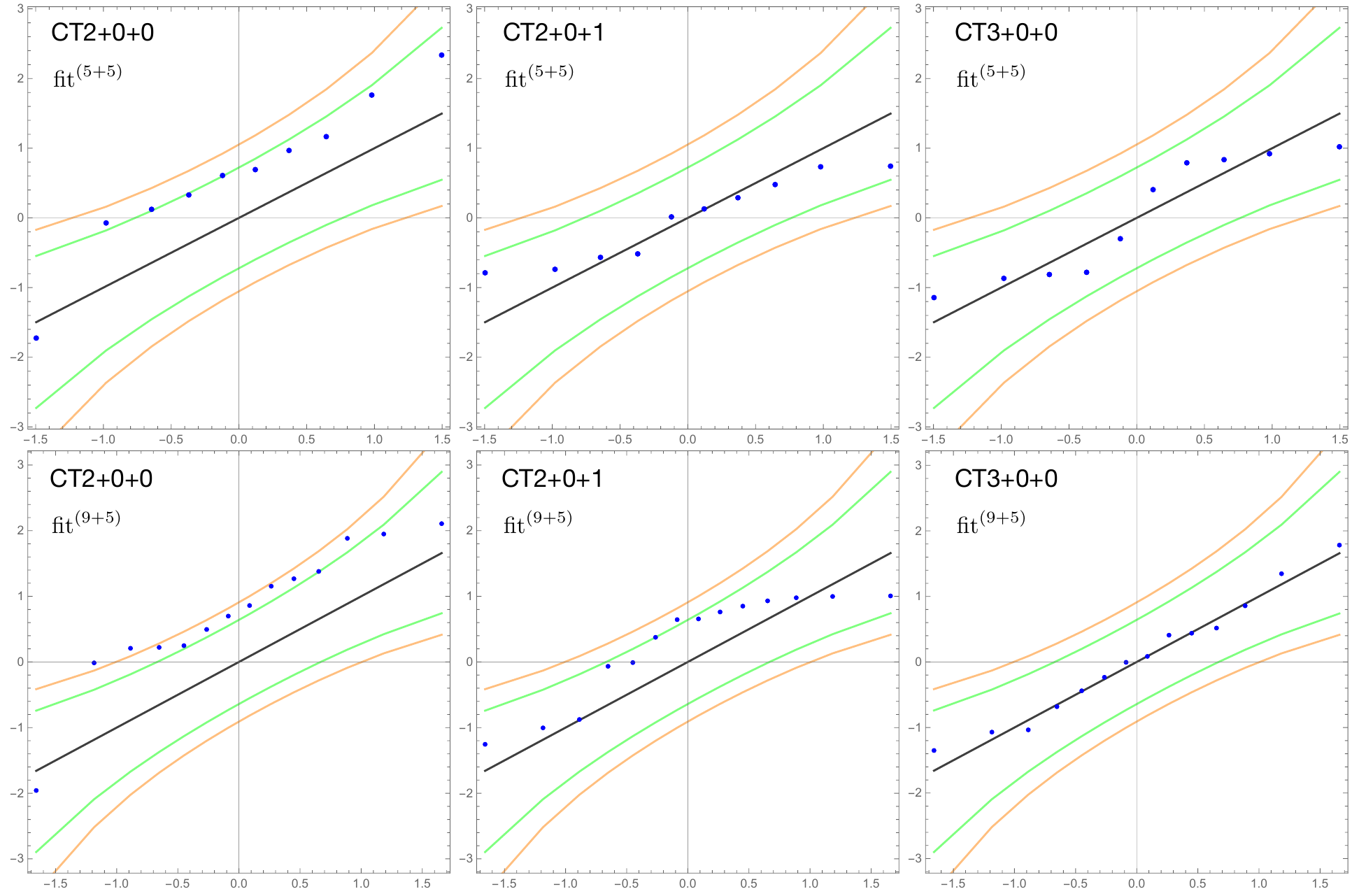}}    
           \caption{    \label{Fig:QQall} 
           Comparison of QQ plots for EFT calculations with different numbers of contact terms. The left, middle, and right panels correspond to the fits CT2+0+0, CT2+0+1, and CT3+0+0, respectively. The upper panels show fits to the lowest (5+5) lattice data points in the $A_1^+[0]$ and $T_1^+[0]$ irreps, while the lower panels display fits to all (9+5) lattice data points. A straight 45$^\circ$ reference line is included to guide the eye. The regions between the green (orange) lines correspond to the 68\% (95\%) confidence intervals.
}
\end{figure*}

In this appendix, we test the statistical consistency of the fits discussed in the main text.
A statistically consistent $\chi^2$-fit should yield residuals that follow a normal distribution.
For each  fit, we present a quantile-quantile (QQ) plot, which allows us to visually assess the difference between a normal distribution and an actual distribution of the residuals in the fit.

Figures~\ref{Fig:QQ4+4}–\ref{Fig:QQall} display QQ plots comparing the quantiles of a standard normal distribution ($x$-axis) to the normalized residuals ($y$-axis), defined as: 
\bea
\text{Res}_i = \frac{ E^{\text{EFT}}(i)
 - E^{\rm LQCD}(i)
}{\Delta E^{\rm LQCD}(i)},
\eea
where $E^{\rm EFT}$ and $E^{\rm LQCD}$ are the finite-volume energy levels from EFT and the lattice, respectively, and $\Delta E^{\rm LQCD}$ denotes the corresponding lattice statistical uncertainties. If the distribution of residuals is close to normal, the points in the QQ plot should lie approximately along the  line $x=y$.
To quantify deviations from normality,  all QQ plots include tail-sensitive confidence bands computed according to Ref.~\cite{Aldor-Noiman01112013}.
The bands are computed such that 68\% (95\%) of pulls from a true standard normal distribution fall inside the corresponding band.
If any point in the QQ plot lies outside the corresponding band, the probability that the residuals follow a normal distribution is less than 32\% (5\%).

As expected, the QQ plot in Fig.~\ref{Fig:QQ4+4} shows good agreement between the sample data and the theoretical distribution from the fit$^{(4+4)}$ of EFT1 (see  the lower panel in Fig.~\ref{Fig_EFT1} for the results of this fit). All points lie well within the 68\% confidence interval. There are no outliers within this band, and any minor deviations suggest only slight differences in the distribution tails.

The situation changes when data points beyond the fourth one are included while retaining the same theory with only  two diagonal contact terms (CT2+0+0). Results from these cases are shown in Fig.~\ref{Fig:QQall}, which includes the five lowest data points from each of the $J=0$ and $J=1$ channels (fit$^{(5+5)}$, the upper left panel), and 9+5 data points (fit$^{(9+5)}$, the lower left panel). In both QQ plots, the residuals deviate significantly from the expected normal distribution: the points curve upward away from the $x=y$ line, indicating pronounced right-skewness (positive skew) in the residuals. In addition, multiple residuals fall above the 68\% confidence band, pointing to statistically significant outliers or extreme values that the theory does not account for. 

A slightly improved, yet still not fully satisfactory, behavior is observed for the fit CT2+0+1, in which an off-diagonal momentum-dependent 
${\cal O}(Q^2)$ contact term is included in addition to the two diagonal contact terms -- see  Fig.~\ref{Fig:QQall} (middle panel).

On the other hand, this picture improves considerably when a third contact term of the form $p^4 + p'^4$ is included. In this case, the QQ plots closely follow the $x=y$ reference line, with all data points residing within the 68\% confidence band -- see Figs.~\ref{Fig:QQall} (right panel). This supports the hypothesis that the residuals follow a normal distribution with mean 0 and standard deviation 1, and confirms that the extended fit with 3 diagonal CTs (CT3+0+0) provides a statistically sound description of the data.

\section{Details of the fits to lattice QCD data and statistical uncertainty estimates} 
\label{appendix_B}

We fit the LECs to FV spectra by minimizing the following $\chi^2$ function,
\begin{equation}\label{chi2_function}
    \chi^2(\{\mathcal{C_{\text{cont}}}\}) = \sum_{i,j}\delta E(i,\{\mathcal{C_{\text{cont}}}\})C^{-1}(i,j)\delta E(j,\{\mathcal{C_{\text{cont}}}\}),
\end{equation}
where $\{\mathcal{C_{\text{cont}}}\}$ is a set of contact terms, $C(i,j)$ is the covariance matrix, and 
\begin{equation*}
  \delta E(i,\{\mathcal{C_{\text{cont}}}\}) = E^{\text{LQCD}}(i)-E^{\text{EFT}}(i,\{\mathcal{C_{\text{cont}}}\})
\end{equation*}
is the difference between the FV energy levels from lattice QCD simulations  
and our fit result  --- see Appendix A in Ref. \cite{Meng:2023bmz} for more details on the general fitting procedure. 

Correlated uncertainties from the original data set are propagated into our calculations using the bootstrap procedure --- see, e.g., Ref.~\cite{Efron:1986hys} for details on the general method. Here, we assume that the original lattice data are multivariate-normally distributed and perform a Monte Carlo simulation to randomly generate numerous additional data sets. Subsequently, we fit each of these generated data sets individually by minimizing the $\chi^2$ function (\ref{chi2_function}), resulting in a distribution of contact terms. We then calculate observables for each set of contact terms within the distribution, e.g., pole positions or effective range parameters.

\section{Consistency checks and estimates of systematic uncertainty}
\label{App:EFT2_results}

\begin{figure*}[h]
    \centering
     \includegraphics[width=0.4\linewidth]
{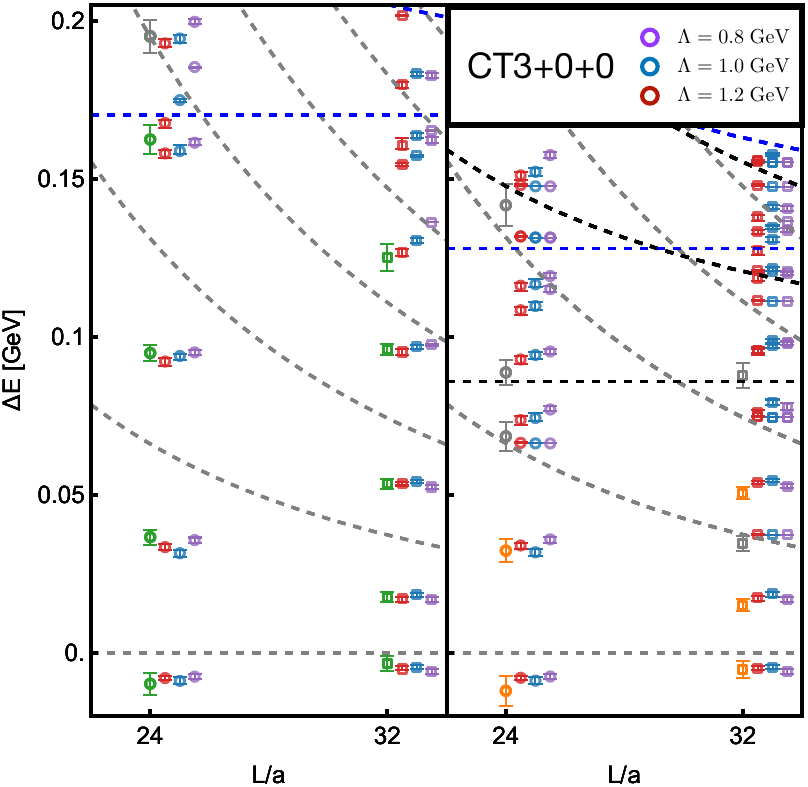}
    \includegraphics[width=0.55\linewidth]{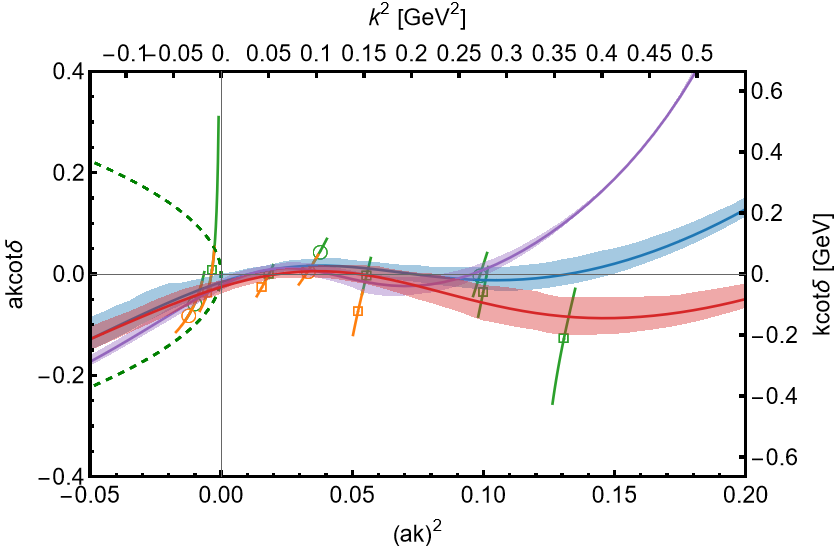}\\
    \caption{Cutoff dependence of the fit CT3+0+0 for the cutoffs $\Lambda = 0.8$ (violet),\ 1.0 (blue), and 1.2 GeV (red), respectively. The left panel shows the energy levels, while the right panel displays $p \cot \delta$ in the lowest channels. The results are shown
including the 1$\sigma$ statistical uncertainty. 
See the caption of Fig.~\ref{Fig_EFT1} for further details.}
    \label{fig:3CT_cutoff}
\end{figure*}

In this Appendix, we provide additional details on the results obtained from fitting all available lattice data within the EFT2 framework, as discussed in Sec.~\ref{Sec:EFT2}.  In particular, we analyze the sensitivity of the results to the choice of the cutoff and  estimate the size of HQSS-violating corrections, which are then propagated to the associated uncertainty in the pole positions.
 
We consider the results of the fits CT2+0+0 and CT3+0+0 with purely diagonal contact interactions for the cutoffs $\Lambda = 0.8,\ 1.0,$ and $1.2$ GeV. The fit CT2+0+0 exhibits a pronounced cutoff dependence  already at very low energies. Due to unphysical poles that emerge close to and below the threshold,  $k \cot \delta$ is significantly distorted in the low-energy region. In contrast, the calculations with three diagonal contact terms remain largely cutoff-independent over the relevant momentum range, from below threshold up to $k \sim 450$ MeV ($k^2 \sim 0.2$ GeV$^2$). This feature is  illustrated in Fig.~\ref{fig:3CT_cutoff}, where the results for all three cutoffs are shown together for the fit CT3+0+0.

\begin{figure*}
\centering
     \includegraphics[width=0.45\linewidth]{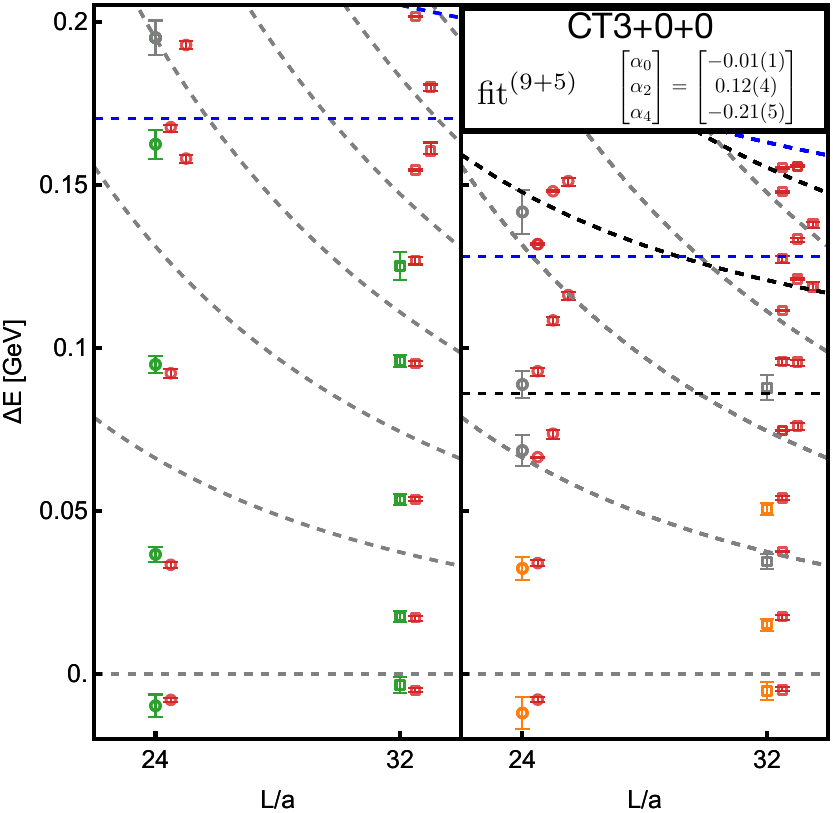}
           \includegraphics[width=0.45\linewidth]
    {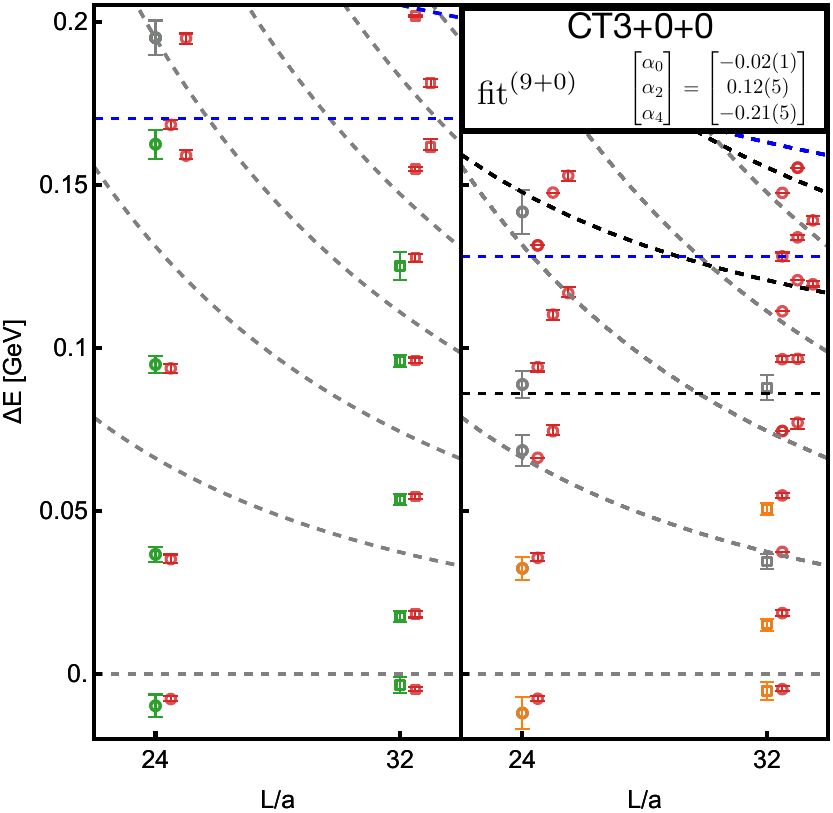}
       \caption{HQSS consistency check: Comparison of the  CT3+0+0 results  from a combined fit to $A_1^+[0]$ and $T_1^+[0]$ irreps (fit$^{(9+5)}$, left panel) with those from a fit to the $A_1^+[0]$ ($J=0$) data only (fit$^{(9+0)}$, right panel). The consistency of the energy levels and the extracted contact terms demonstrates very good agreement with HQSS.
}
    \label{fig95_vs_90}
\end{figure*}

To estimate the impact of HQSS violation, we compare the fit to the $J=0$ channel data in $A_1^+[0]$ (fit $^{(9+0)}$) with the fit to the full lattice dataset involving both the $A_1^+[0]$ and $T_1^+[0]$ irreps (fit $^{(9+5)}$), as shown in Fig.~\ref{fig95_vs_90}. As expected, the calculated energy levels and the extracted contact terms are fully consistent between the two fits within statistical uncertainties -- see the results in the left and right panels.
Consequently, following the strategy adopted in EFT1, we allow for a $10\%$ HQSS violation in ${\cal C}_0$ for the fit$^{(9+5)}$. This introduces an additional uncertainty, which propagates to the pole positions reported in Table~\ref{Table:EFT1_EFT2_results}. The estimated HQSS-violation uncertainty is smaller than the corresponding statistical error.

 \end{document}